\documentclass[aps,showpacs,pre,twocolumn]{revtex4}

\usepackage{psfig}
\psfigurepath{figs}

\def \xpn{\mu}
\def \invdens{\rho}

\newcommand{\Eqn}[1]{Eq.~(\ref{#1})}     
\newcommand{\Sec}[1]{Section~\ref{#1}}     
\newcommand{\Apd}[1]{Appendix~\ref{#1}}     
\newcommand{\Fig}[1]{Fig.~\ref{#1}}

\begin{document}
\title{Shortest paths on systems with power-law distributed long-range
  connections}
\author{C.~F.~Moukarzel$^{1,2}$~\footnote{Corresponding author:
    cristian@mda.cinvestav.mx}}
\affiliation{$^1$ Depto. de F\'\i sica Aplicada, CINVESTAV del IPN,\\
  Av.~Tecnol\'ogico Km 6, 97310 M\'erida, Yucat\'an, M\'exico }
\author{M.~Argollo de Menezes}
\affiliation{$^2$ Instituto de F\'{\i}sica, Universidade Federal Fluminense\\
  Av.~Litor\^anea s/n, 24210-340, Niter\'oi, RJ, Brazil}
\date{\today}
\begin{abstract}
  We discuss shortest-path lengths $\ell(r)$ on periodic rings of size $L$
  supplemented with an average of $pL$ randomly located long-range links whose
  lengths are distributed according to $P_l \sim l^{-\xpn}$. Using rescaling
  arguments and numerical simulation on systems of up to $10^7$ sites, we show
  that a characteristic length $\xi$ exists such that $\ell(r) \sim r$ for
  $r<\xi$ but $\ell(r) \sim r^{\theta_s(\xpn)}$ for $r>>\xi$. For small $p$ we
  find that the shortest-path length satisfies the scaling relation
  $\ell(r,\xpn,p)/\xi = f(\xpn,r/\xi)$.  Three regions with different
  asymptotic behaviors are found, respectively: a) $\xpn>2$ where
  $\theta_s=1$, b) $1<\xpn<2$ where $0<\theta_s(\xpn)<1/2$ and, c) $\xpn<1$
  where $\ell(r)$ behaves logarithmically, i.e. $\theta_s=0$. The
  characteristic length $\xi$ is of the form $\xi \sim p^{-\nu}$ with
  $\nu=1/(2-\xpn)$ in region b), but depends on $L$ as well in region c). A
  directed model of shortest-paths is solved and compared with numerical
  results.
\end{abstract}
\maketitle 
\section{Introduction}
It has been known for long that slowly decaying long-ranged (LR) interactions
can drastically change the critical behavior of a system.  A well studied
example is the one-dimensional Ising model with $J(r) \sim r^{-\xpn}$
\cite{TL69,DI71,FMNC72,KP76,MB92,CO95,LBF96,LBC97,MO98,BDE99}, which is
relevant for the Kondo problem~\cite{HF69,AYE69} among others. If $\xpn>2$
there is no ordered phase at any finite temperature, the same as if only
short-ranged interactions were present.  When $\xpn=2$ the magnetization
undergoes a finite jump at $T_c>0$, while all derivatives of the free energy
remain finite (essential singularity).  When $\xpn<2$ the model displays a
second-order phase transition with $\xpn$-dependent critical indices, which
take their classical, or Mean-Field (MF) values for $\xpn< 1.5$. On approach
to $\xpn=2$ from below, the correlation-length exponent diverges, signaling
the appearance of an essential singularity. This divergence is of the form
$\nu \sim (2-\xpn)^{-1/2}$~\cite{KP76} for Ising and $(2-\xpn)^{-1}$ for
$n$-component models with $n>1$ (but see~\cite{CMT97}, where $\nu \sim
(2-\xpn)^{-1} \forall n$ is suggested).  A comprehensive account of what is
known for Ising systems with LR interactions has been given by Luijten and
Bl\"ote~\cite{LBC97}.
\\
For $d$-dimensional $n$-component systems with ferromagnetic interactions
decaying as $1/r^{d+\sigma}$, Fisher, Ma and Nickel~\cite{FNMC72} propose that
the lower critical decay rate is given by $\sigma=d/2$, or equivalently that
the upper critical dimension is $d_u=2\sigma$. For $\sigma < d/2$ the critical
indices take their MF values, for $d/2 <\sigma<2$ they are $\sigma$-dependent,
and for $\sigma>2$ they take their short-range (SR) values.  Similar
investigations have been conducted for Potts~\cite{GUC93,CMT97,ML99,BDDP99},
Heisenberg\cite{WSE01,RSP00,FQA98,RC96b,RC92,FC88}, and
other~\cite{TAR00,RIR99} models.
\\
The following picture is often found: for small enough decay rate $\xpn$, MF
indices are obtained. Upon increasing $\xpn$ a regime follows where critical
indices change continuously with $\xpn$ until finally SR indices are
recovered. In a loose sense one can say that the addition of LR interactions
changes the ``effective dimension'' of the system, although in a way that may
depend on the specific model considered.  This idea has been exploited to
study the scaling behavior of critical systems above their effective upper
critical dimension $d_u$, while still working on lattices of low Euclidean
dimension~\cite{LBF96}.  The connection between LR interactions and
dimensionality was also briefly touched upon by Scalettar~\cite{SC91}. A
possible way to define an effective dimension, which is in general
model-dependent, is to do so through the hyperscaling relation
$(2-\alpha)=d\nu$, as~\cite{BDE99,BDDP99} $d_{eff}=\nu^{-1}(2 - \alpha)$.
\\
An alternative paradigm for the problem of LR interactions considers systems
on a $d$-dimensional lattice supplemented with randomly distributed LR bonds
of unit strength, which are present with probability $p_{ij} \sim
r_{ij}^{-\xpn}$. Notice that in this case the system has disorder: it is the
probability for a given bond to be present, and not its strength, what decays
with distance. These two ways to introduce LR interactions; decaying strength
(DS) and decaying probability (DP), are not in principle equivalent.  It is
well known that disorder may change the critical behavior if the specific-heat
exponent $\alpha$ is negative. The DP paradigm is on the other hand relevant
for a number of problems in which connectivity, and not the strength of the
interaction, is determinant of the physical behavior.  Examples of problems of
this kind are the magnetic~\cite{CMSI85} and conductive~\cite{CCR85,XCED96}
properties of polymeric chains, where the probability of crosslinks between
two monomers decays as a power-law of the chemical distance between them,
conduction in insulating matrices with one dimensional conducting
inclusions~\cite{BSTC97} whose length distribution is ``broad'', neural
networks~\cite{YCC99,LHCF00}, geodesic propagation on spaces with topological
singularities (wormholes), the spread of fire or diseases~\cite{MS99,PVE01b},
etc.
\\
Networks built according to the DP paradigm of LR interactions may be
characterized entirely in geometrical (or topological) terms, because all
bonds have the same strength. Thus it appears for example possible to define
the relationship between effective dimension $d_{eff}$ and decay rate $\xpn$
of interactions in purely geometric terms for these networks.
\\
A useful topological characterization of random networks is the \emph{Graph
  Dimension} $d_g$, defined as follows: if $V(\ell)$ is the average number of
sites that can be reached from a given one in $\ell$ steps between connected
neighbors, then $V(\ell) \sim \ell^{d_g}$ asymptotically.  We now let
$\ell(r)$ be the average smallest number of links needed to join two points
separated by an Euclidean distance $r$ (the ``shortest-path length''), which
behaves asymptotically as $r^{\theta_s}$, where $\theta_s$ is the
shortest-path dimension~\cite{BHF96}. Since $V(r) \sim r^d$, the above
relations imply that $d_g = d/\theta_s$, and we see that the asymptotic
behavior of $\ell(r)$ defines the graph dimension $d_g$.
\\
In this work we study shortest-paths on DP networks, i.e.  $d$-dimensional
lattices with the addition of an average of $p$ LR bonds (or \emph{shortcuts})
per site, whose length is distributed according to $P_l \sim l^{-\xpn}$. We
shall concentrate mostly on the case $d=1$, where numerical simulations are
easiest.  DP networks with power-law distributed LR bonds have been recently
considered in one dimension both from the point of view of Random Walk
properties\cite{JBS00} and Shortest-Path lengths~\cite{SCS01}, but for small
system sizes. We will later discuss some of the conclusions in~\cite{SCS01},
which appear to need revision in the light of our results.
\\
In \Sec{sec:model} several definitions which are relevant for our problem of
shortest-paths on 1d DP networks are given.  Simple rescaling arguments are
used in \Sec{sec:rescaling} to show that $\xpn=2d$ is a critical decay rate,
such that for $\xpn>2d$, LR bonds are unimportant on large scales. For
$\xpn<2d$ on the other hand, when $p$ is small these arguments predict the
existence of a characteristic length $\xi \sim p^{-1/(2-\xpn)}$, beyond which
LR bonds are important.  In \Sec{sec:naive} a directed model is introduced for
shortest-paths in 1d, which turns out to be exact for $\xpn>2$ and still
provides an useful upper bound when $\xpn <2$.  In \Sec{sec:numerical} our
extensive numerical results for shortest-path lengths $\ell(r)$ in one
dimension are described and compared to theoretical predictions. Finally,
\Sec{sec:conclusions} contains a discussion of our results.
\section{DP networks and rescaling}
\label{sec:model}
We start with an arbitrary $d$-dimensional lattice made up of $N=L^d$ sites,
and its corresponding SR bonds. In addition to these, DP networks are defined
to have an average of $pL^d$ LR bonds, or \emph{shortcuts}, whose lengths and
locations are random. This is done in practice by letting one LR bond stem
from each site $i$ with probability $p$. The neighbor $j$ at the other end of
each LR bond is randomly chosen with a probability $P(j|i)$, that is a
decaying function of the Euclidean distance $r_{ij}=|\vec x_i-\vec x_j|$
between sites $i$ and $j$.
\\
For a given realization of shortcuts, the shortest-path length $\ell_{ij}$ is
defined as the minimum number of connected-neighbor steps needed to join sites
$i$ and $j$. This quantity is measured as a function of Euclidean distance
$r_{ij}$, and averaged over disorder (realizations of shortcuts).  After
disorder average, $\ell(r)$ is the average ``cost'' of joining two points
separated by an Euclidean distance $r$, and is defined as
\begin{equation}
\ell(r) = \sum_{ij} <\ell_{ij}> \delta(r_{ij}-r) /\sum_{ij} \delta(r_{ij}-r),
\end{equation}
where $<>$ means disorder average.  
\subsection{Rescaling}
\label{sec:rescaling}
Consider now dividing the $d$-dimensional lattice into ``blocks'' of linear
dimension $b$, such that $1<<b<<L$, and regard two sites $I$ and $J$ of this
new lattice to be connected by a shortcut if \emph{any} pair $\{i \in I, j \in
J\}$ is connected by a shortcut. We allow for at most one shortcut between
rescaled sites since, for the purpose of shortest-paths, the only fact that
matters is whether two sites are connected or not. If the original pairs $ij$
are connected by a shortcut with probability $p_{ij}$, the rescaled
probability $\tilde q_{IJ}=1-\tilde p_{IJ}$ for blocks $I$ and $J$ not to be
connected is given by
\begin{equation}
\tilde q_{IJ} = \prod_{i \in I, j\in J} (1-p_{ij})= \prod_{i \in I, j\in J}
q_{ij},
\label{rescaling-def}
\end{equation}
which for large distances $|i-j|>> b$ can be approximated as $ \tilde q_{IJ} =
q_{ij}^{b^{2d}} $. This can be written as $\tilde q(r/b) = \left ( q(r) \right
)^{b^{2d}}$ and therefore $\lambda(r) = \log q(r)$ transforms in a simple way
under rescaling,
\begin{equation}
\label{eq:rescaling}
\tilde \lambda(r/b) =  b^{2d} \lambda(r).
\end{equation}
Thus
\begin{equation}
\label{eq:invariant-distribution}
p(r)=\left(1-e^{-\invdens/r^{\xpn}}\right),
\end{equation}
retains its functional form under rescaling, i.e.
\begin{equation}
\tilde p(r)=\left(1-e^{-\tilde \invdens/r^{\xpn}}\right),
\end{equation}
with $\tilde \invdens = b^{(2d-\xpn)} \invdens$. The condition that the system
contains a total of $pL^d$ LR bonds is ensured by imposing
\begin{equation}
p = S_d \int_1^L p(r) r^{d-1} dr,
\label{eq:normalization}
\end{equation}
where $S_d$ is the surface of a $d$-dimensional hypersphere of radius one.
This relationship fixes $\invdens$ as a function of $p$ and $L$. In the limit
of small $p$, $\invdens$ turns out to be proportional to $p$. Notice that,
because of the multiplicative rescaling \Eqn{rescaling-def}, a pure power law
is not strictly invariant under rescaling. But the true invariant distribution
\Eqn{eq:invariant-distribution} can be very well approximated by a power law
for large distances $r$ such that $\invdens/r^\xpn <<1$.  Restricting
ourselves to the limit of small $\invdens$ (or $p$) we can thus work with a
power-law distribution of shortcut lengths. In the following we consider
\begin{equation}
p(r) = C \frac{p}{r^{\xpn}},
\label{eq:pl-approx}
\end{equation}
where the normalization constant $C$ is chosen so as to satisfy
\Eqn{eq:normalization}. In \Apd{apd:rescaling} we show that $p$ rescales as
\begin{equation}
\label{eq:p-exponent}
\tilde p = b^{y_p} p,
\end{equation}
with
\begin{equation}
y_p = \left \{
\begin{array}{lrr}
d & \hbox{for} & \xpn \leq d \\ \\
2d-\xpn &\qquad \hbox{for} & \xpn > d
\end{array}
\right .
\label{eq:pscalinglaw}
\end{equation}
Notice that expressions similar to \Eqn{eq:p-exponent} and
\Eqn{eq:pscalinglaw} give the renormalized coupling constant of the
one-dimensional LR Ising model at low temperatures~\cite{CO95}.
\\
It follows that $p=0$ is a line of fixed points in the $\xpn,p$ space of
parameters.  For $\xpn<2d$ this fixed line is repulsive, and becomes
attractive for $\xpn>2d$. Thus for $\xpn > 2d$ the density of LR bonds is
renormalized to zero under rescaling, and $\xpn_c=2d$ is the upper critical
decay rate above which LR bonds are irrelevant, and SR behavior is recovered.
\subsection{Naive Paths: An approximate model in one dimension}
\label{sec:naive}
\begin{figure}[htb] 
\centerline{\psfig{figure=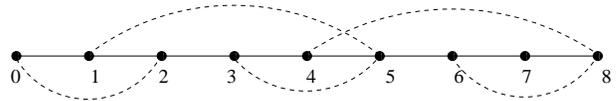,width=8cm,angle=0}} 
\caption{{}Full lines are SR bonds, dashed LR bonds. One possible shortest
  path between $0$ and $8$ is $\{0$-$1,1$-$5,5$-$4,4$-$8\}$ and has length
  four. The naive path uses all rightwards LR bonds available at each site,
  i.e.  $\{0$-$2,2$-$3,3$-$5,5$-$6,6$-$8\}$ and has length five in this
  example}
\label{fig:paths}
\end{figure}

Consider a directed path which starts at $t=0$ from $x_0=0$, proceeds always
to the right, and is built by using at each site any LR bond available,
provided this bond does not take the path further to the right than $r$. We
call the path so defined the ``naive path'' between $0$ and $r$. As compared
with the actual shortest path, this construction neglects the possibility of
turnbacks, or that certain LR bonds may not be used (See \Fig{fig:paths}).  We
will later see that under certain circumstances, the naive-path approximation
gives a reasonable estimate for shortest-path lengths. But even if this is not
the case, the former constitutes an upper bound for the shortest-path length,
and thus still provides useful information.
The naive-path length $\ell_n(r)$ is the number of timesteps it takes to reach
$r$, and can be estimated in the following way. At time $t$ the walker sits at
site $x_t$. From this site, with probability $p$ a LR bond (of random length
$l_t$) stems rightwards. The walker now proceeds along this LR bond, provided
it does not go further to the right than $r$. The joint probability
$\tilde p_t$ that a bond is present at $x_t$, and its length is not larger
than $r-x_t$ is
\begin{equation}
\tilde p_t =  p \sum_{l=1}^{r-x_t} P_l 
\label{eq:tildep}
\end{equation}
Thus at time $t$ the walker goes one unit to the right with probability
$\tilde q_t=1-\tilde p_t$, and $l_t$ units with probability $\tilde p_t$.
Therefore in average
\begin{equation}
x_t = x_{t-1} + 1 + \tilde p_t ( \bar l_t - 1),
\label{eq:discr0}
\end{equation}
where $\bar l_t$ is the average length of a LR bond which is not larger than
$r-x_t$, i.e.
\begin{equation}
\bar l_t = \frac{\sum_{l=1}^{r-x_t} lP_l}{\sum_{l=1}^{r-x_t} P_l}=
\frac{p}{\tilde p}\sum_{l=1}^{r-x_t} lP_l
\label{eq:alen0}
\end{equation}
Thus \Eqn{eq:discr0} reads
\begin{equation}
x_t = x_{t-1} + 1 + p G(r-x_t),
\label{eq:discr1}
\end{equation}
where 
\begin{equation}
G(n)=\sum_{l=1}^{n} (l-1)P_l.
\label{eq:G}
\end{equation}  
Within a continuous-time, continuous-space approximation we put
\begin{equation}
\dot x(t) = 1 + p G\left (r-x(t)\right ),
\label{eq:cont0}
\end{equation}
which shall be solved with boundary conditions $x(t=0)=0$ and
$x(t=\ell_n(r)-1)=r-1$ (notice that Eqns. (\ref{eq:tildep}) and
(\ref{eq:alen0}) are only defined for $x_t \leq r-1$). This can be formally
integrated to give
\begin{equation}
\ell_n(r)=1 + \int_1^r \frac{dx}{1+ p G(x)}.
\label{eq:naive_formal}
\end{equation}
We will analyze this result and compare it with our numerical results in the
following sections.
\section{Numerical results in one dimension}
\label{sec:numerical}
In this section, numerical results are presented for periodic rings of up to
$10^7$ sites. One LR bond stems from each site with probability $p\leq 1$.
Its random length $l$ is obtained by first generating a real random variable
$z$ such that $1\leq z<(L/2+1)$ with $P(z) \sim z^{-\xpn}$, and then taking
its integer part: $l=Int(z)$.  Lattice sizes are $L_k=10^{3+k/2}$ for
$k=0,1,\cdots,8$. The density of LR bonds is $p=0.001, 0.003, 0.01, 0.033,
0.1$ and $1.0$.  Shortest paths are identified by Breadth-First-Search
(BFS)~\cite{LH90,CLRI93}, and averages are taken over $10^4$ samples.
Altogether the results presented in this work involve an amount of
computational work equivalent to approximately $10^{12}$ sites.
\begin{figure}[h]
\vbox{
\leftline{\hskip -1cm
\psfig{figure=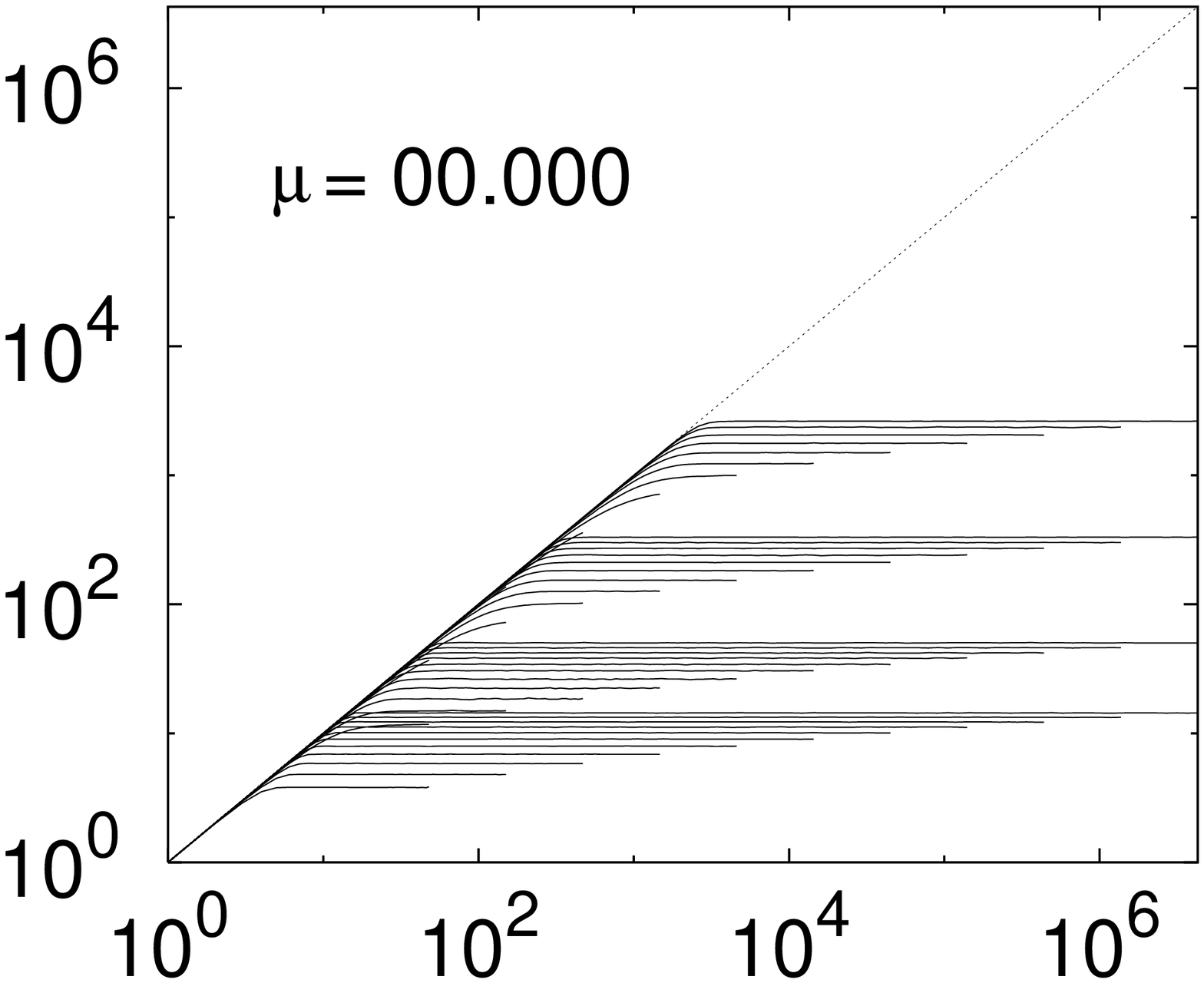,width=5.3cm,angle=270} \hskip -1cm
\psfig{figure=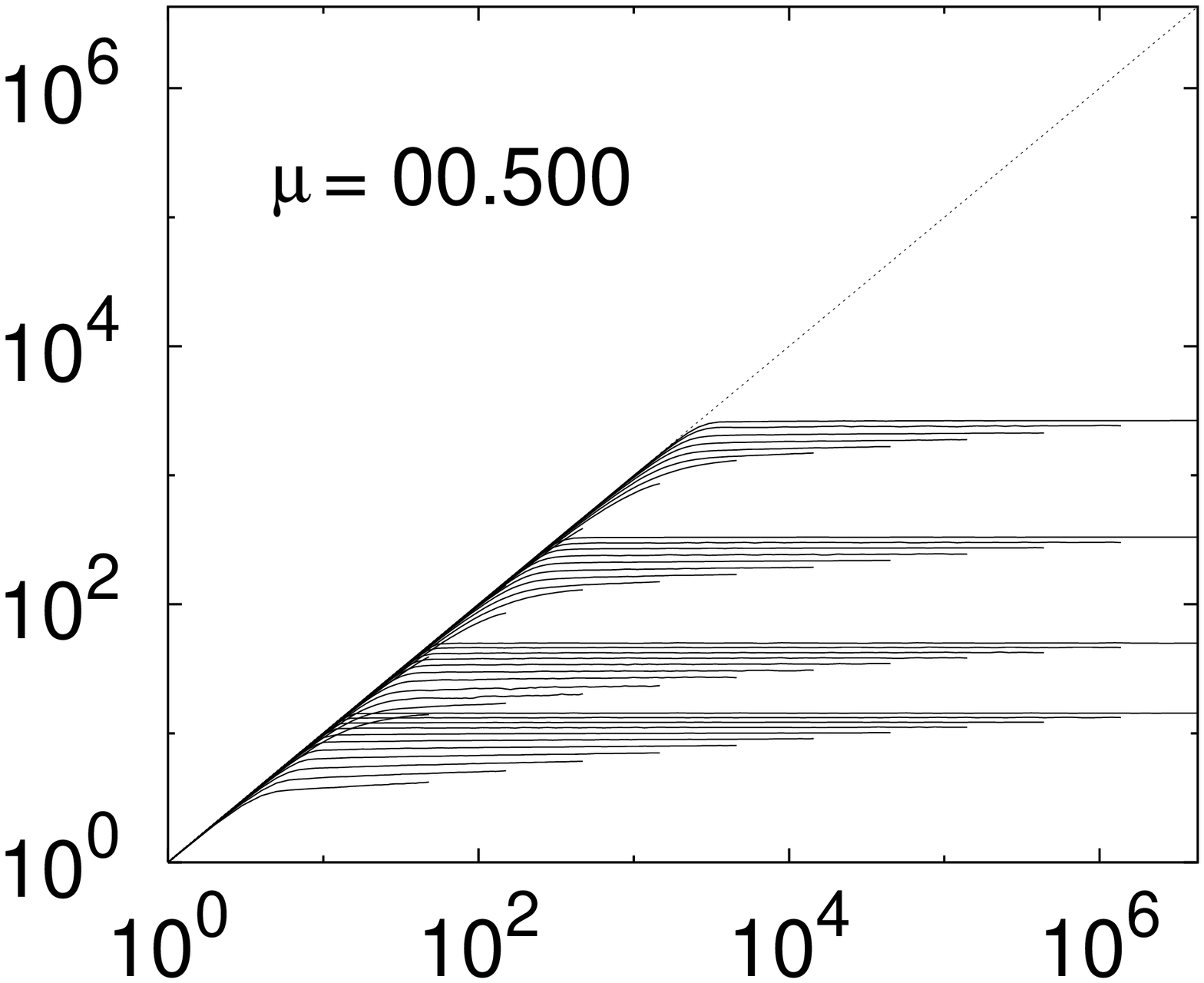,width=5.3cm,angle=270}
}\leftline{\hskip -1cm
\psfig{figure=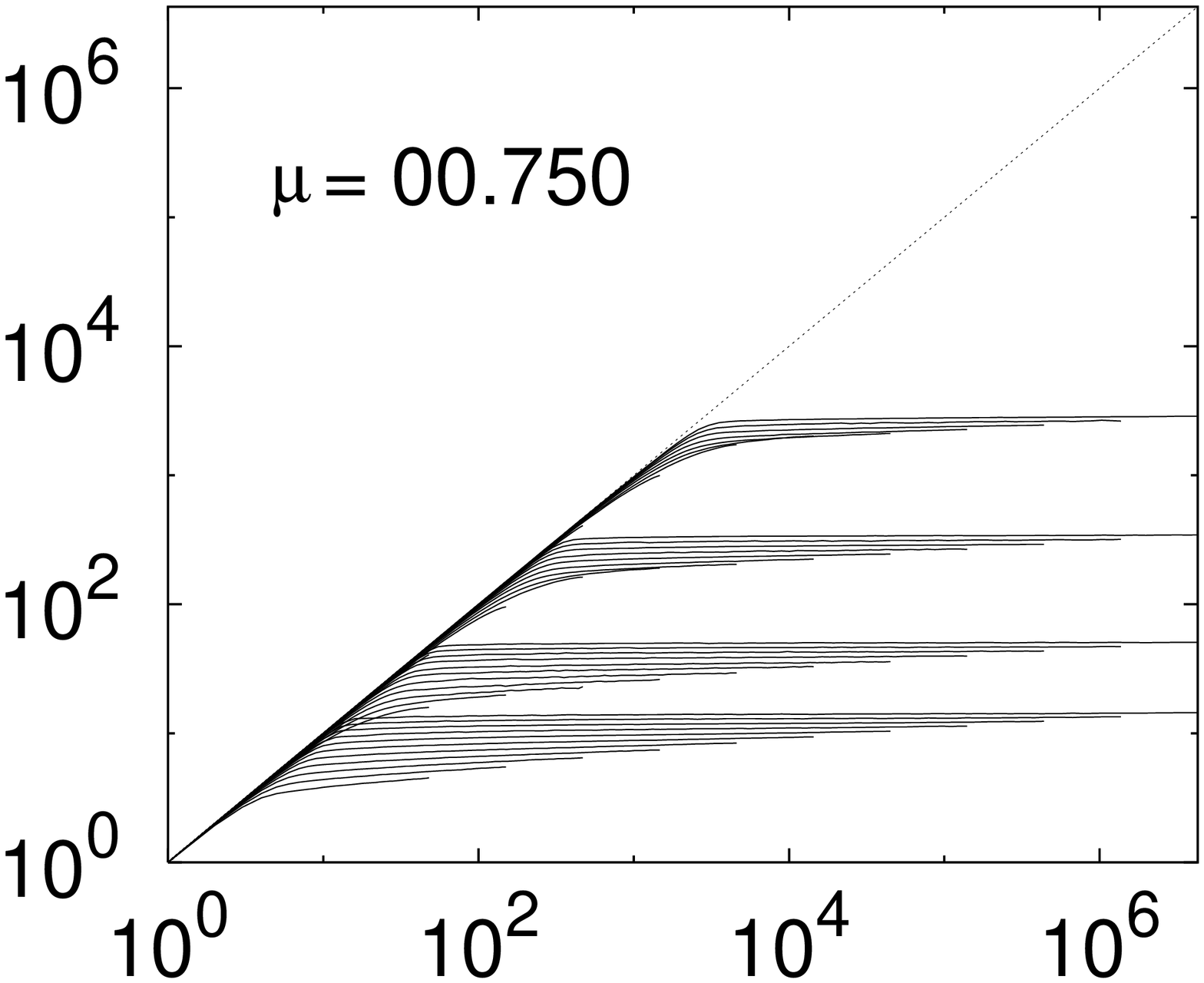,width=5.3cm,angle=270} \hskip -1cm
\psfig{figure=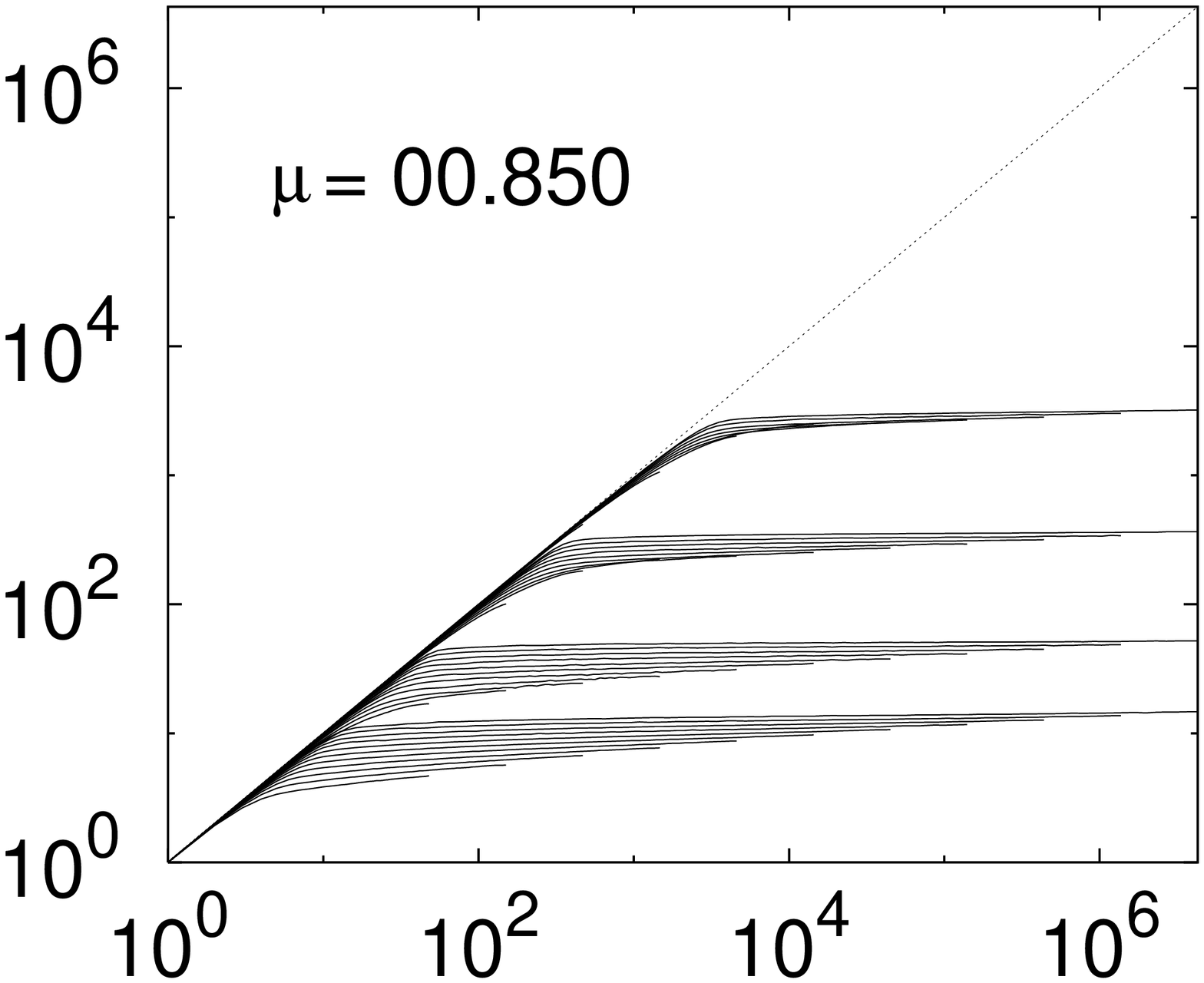,width=5.3cm,angle=270}
}\leftline{\hskip -1cm
\psfig{figure=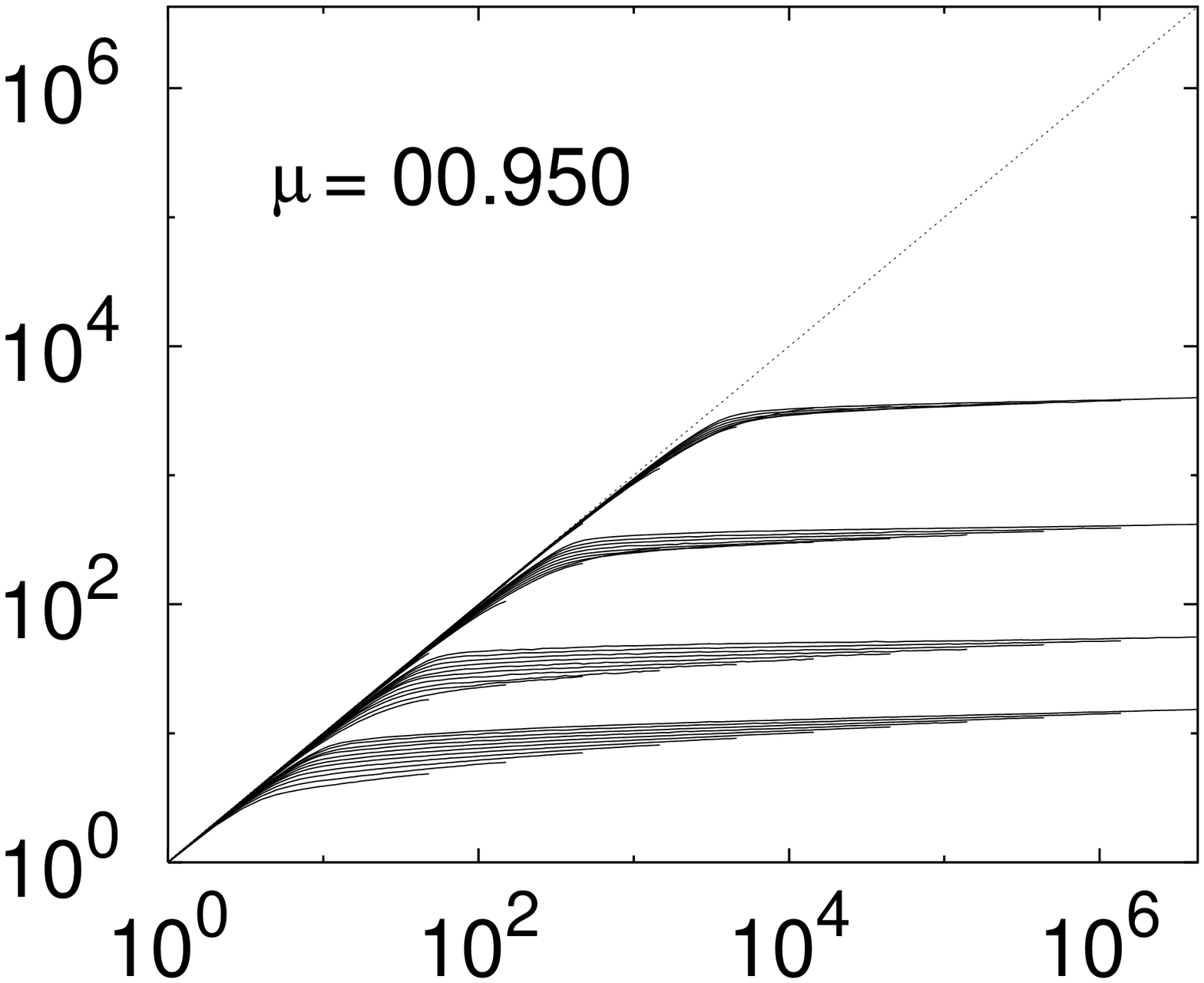,width=5.3cm,angle=270} \hskip -1cm
\psfig{figure=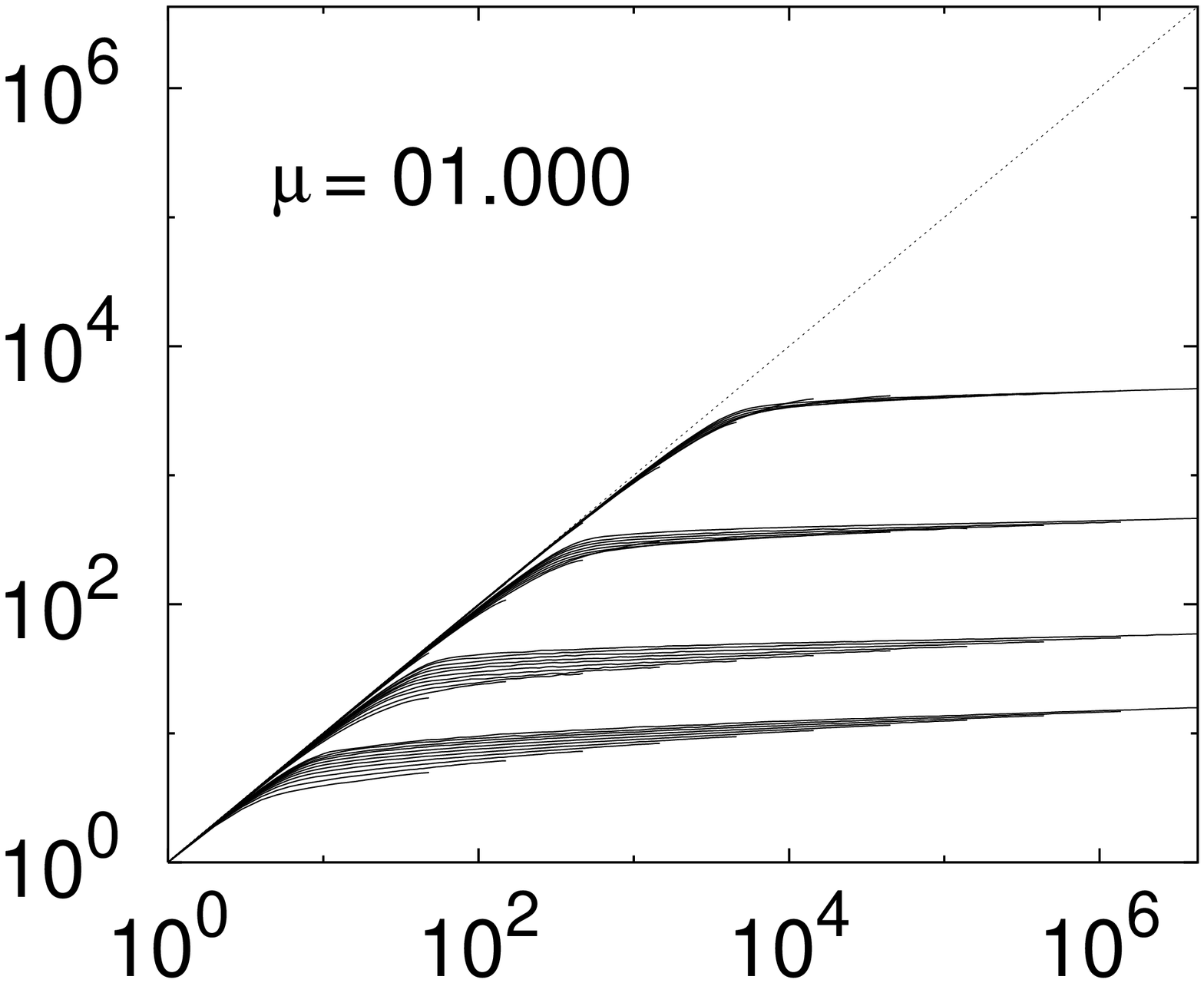,width=5.3cm,angle=270}
}}
\caption{{} Average shortest-path length $\ell(r)$ \emph{vs} $r$. Numerical
  averages (full lines) over $10^4$ samples are shown for systems of size
  $L_k=10^{3+k/2}$ with $k=0,1,\cdots,8$. The dashed line is $\ell(r)=r$. The
  local density $p$ of LR bonds is $p=10^{-3}, 10^{-2},10^{-1}$ and $1$. The
  different cases can be told apart by noticing that larger values of $p$
  result in lower values of $\ell$. }
\label{fig:ell_0a}
\end{figure}
\begin{figure}[h]
\vbox{
\leftline{\hskip -1cm
\psfig{figure=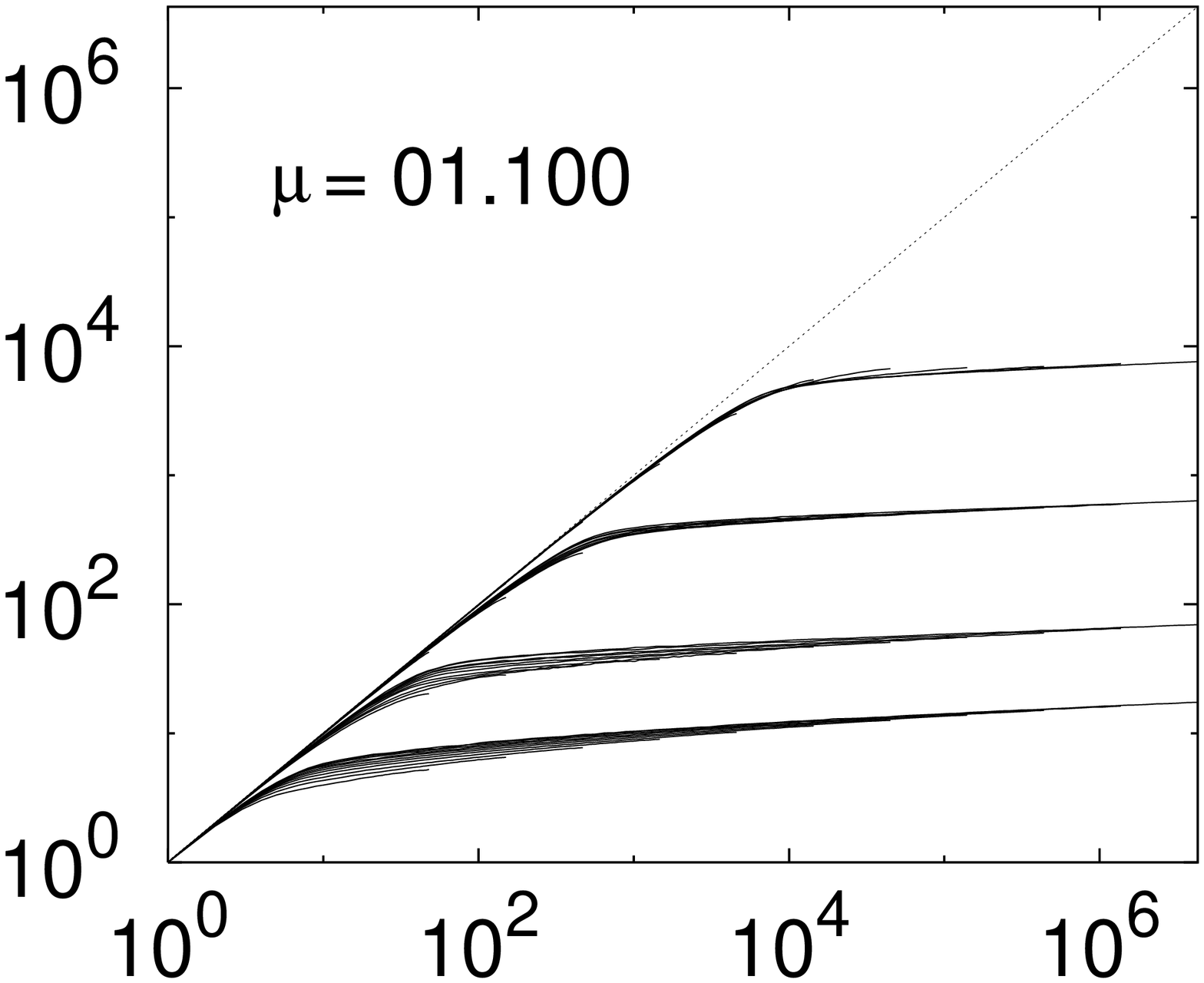,width=5.3cm,angle=270} \hskip -1cm
\psfig{figure=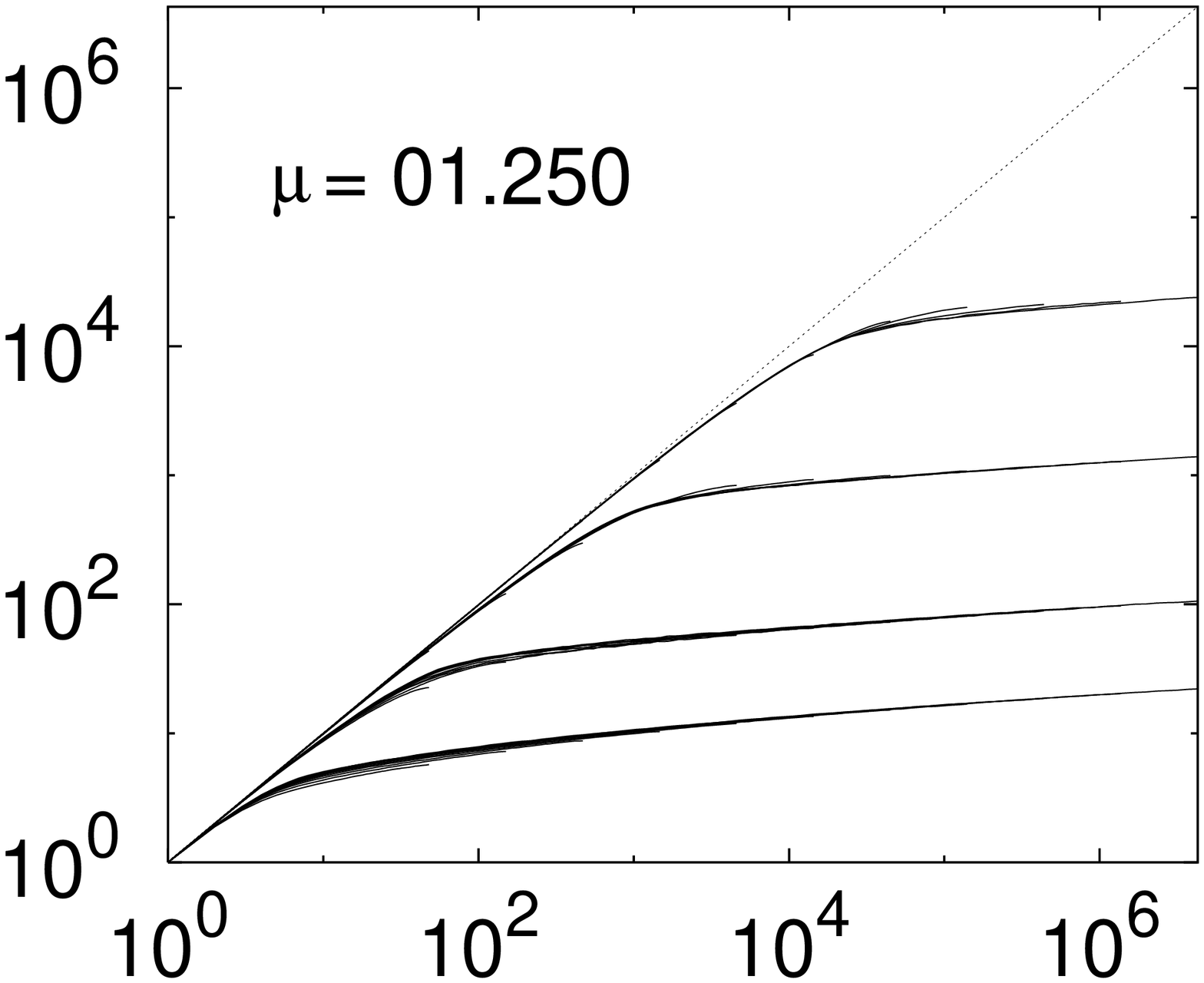,width=5.3cm,angle=270}
}\leftline{\hskip -1cm
\psfig{figure=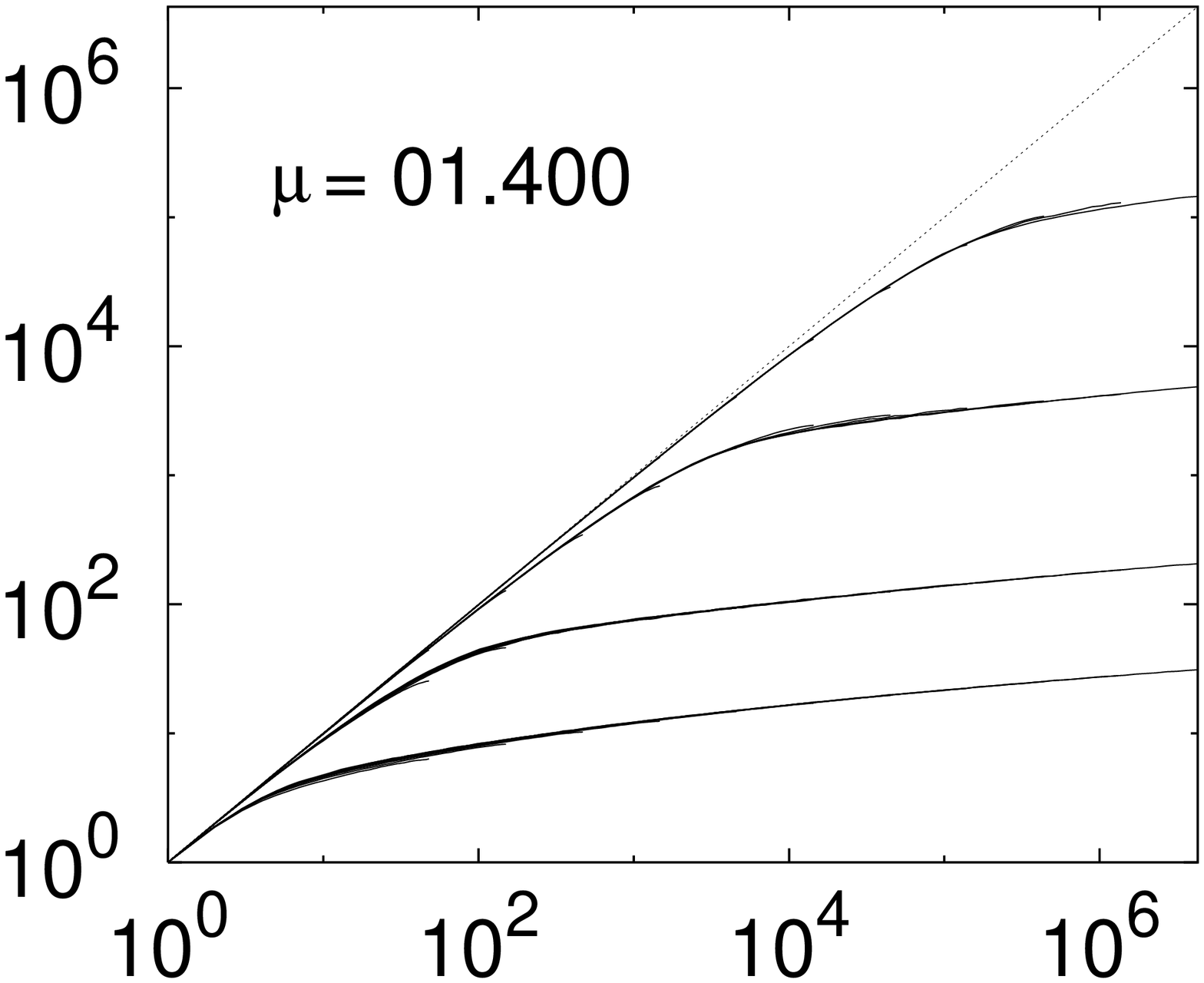,width=5.3cm,angle=270} \hskip -1cm
\psfig{figure=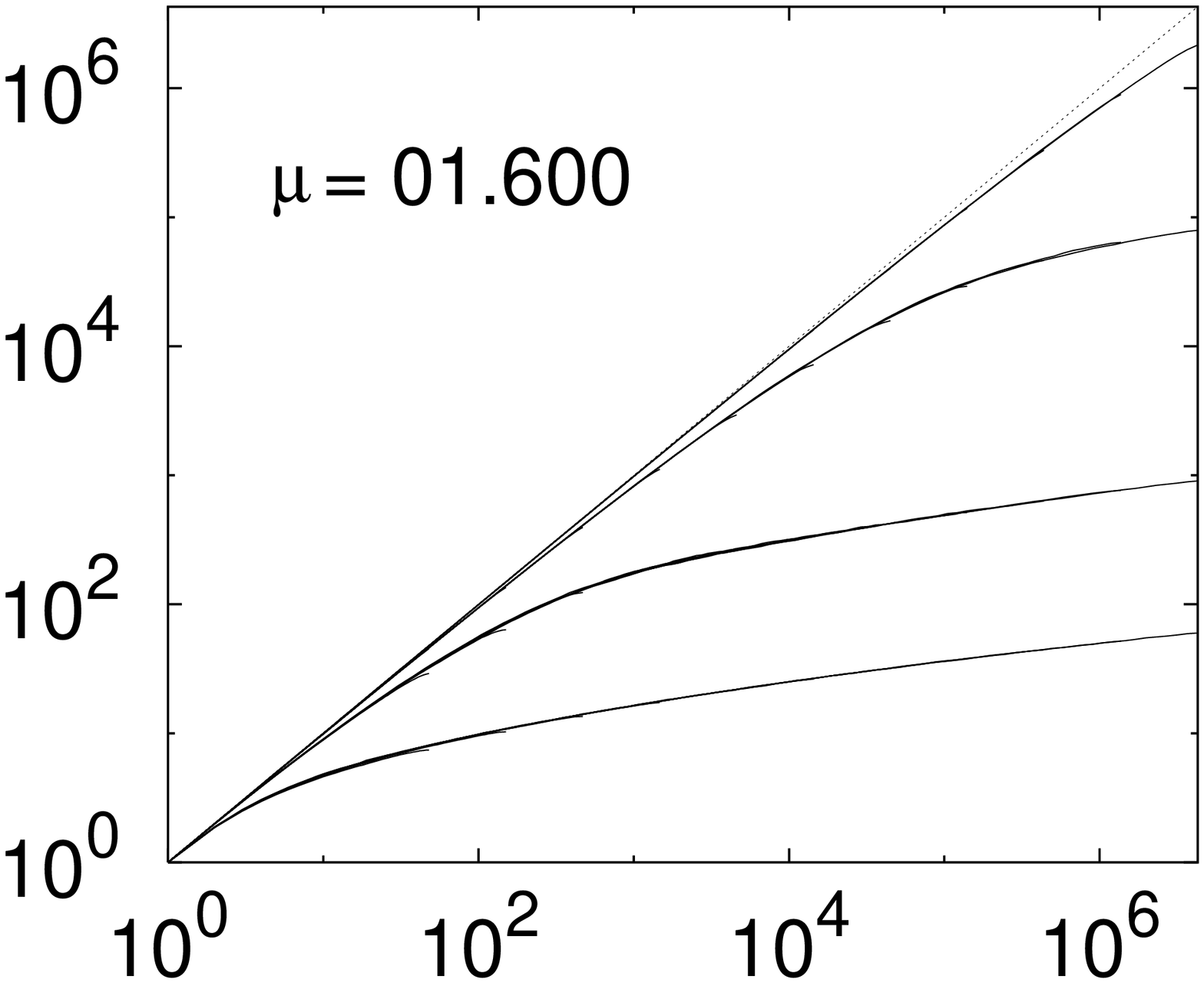,width=5.3cm,angle=270}
}\leftline{\hskip -1cm
\psfig{figure=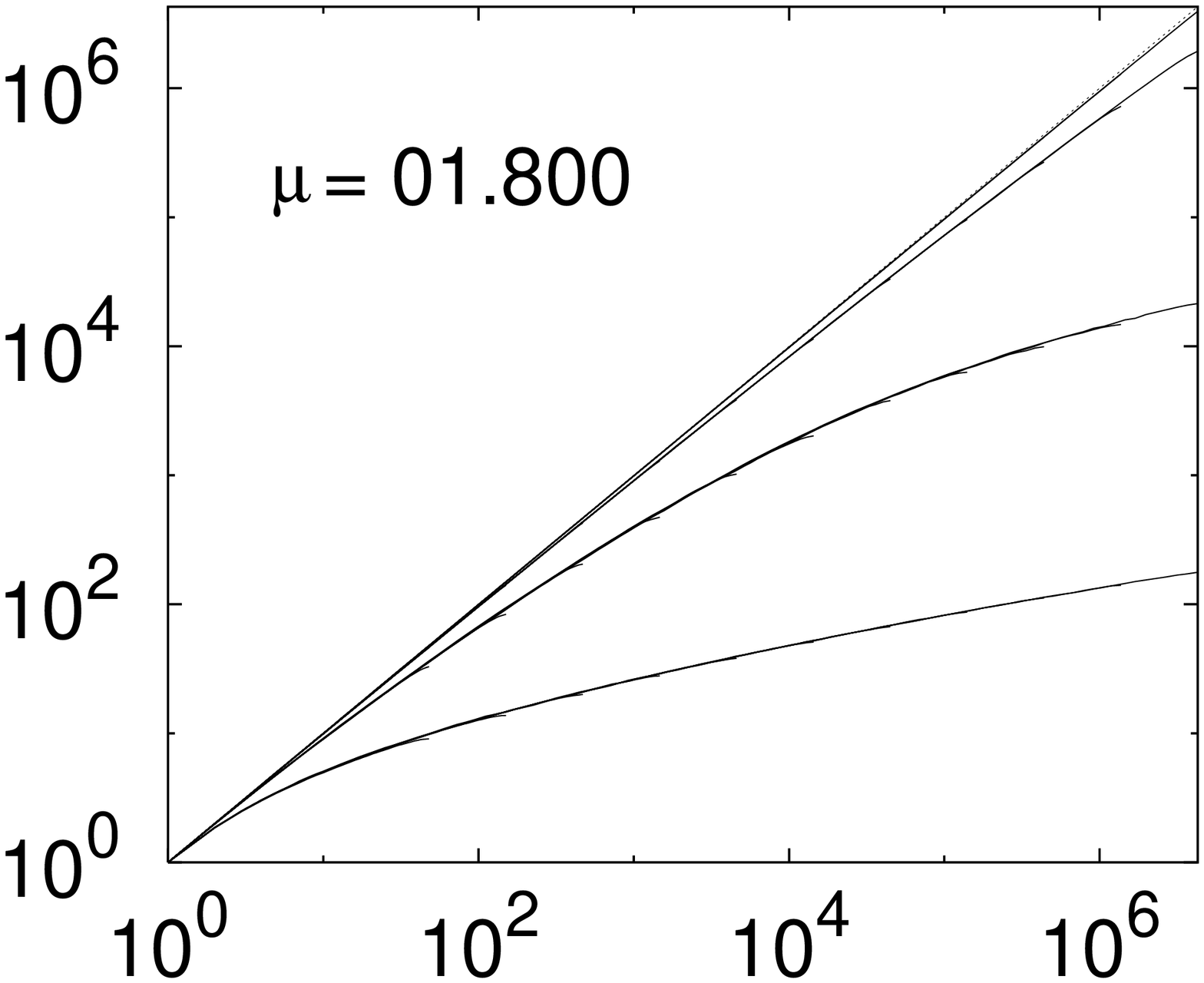,width=5.3cm,angle=270} \hskip -1cm
\psfig{figure=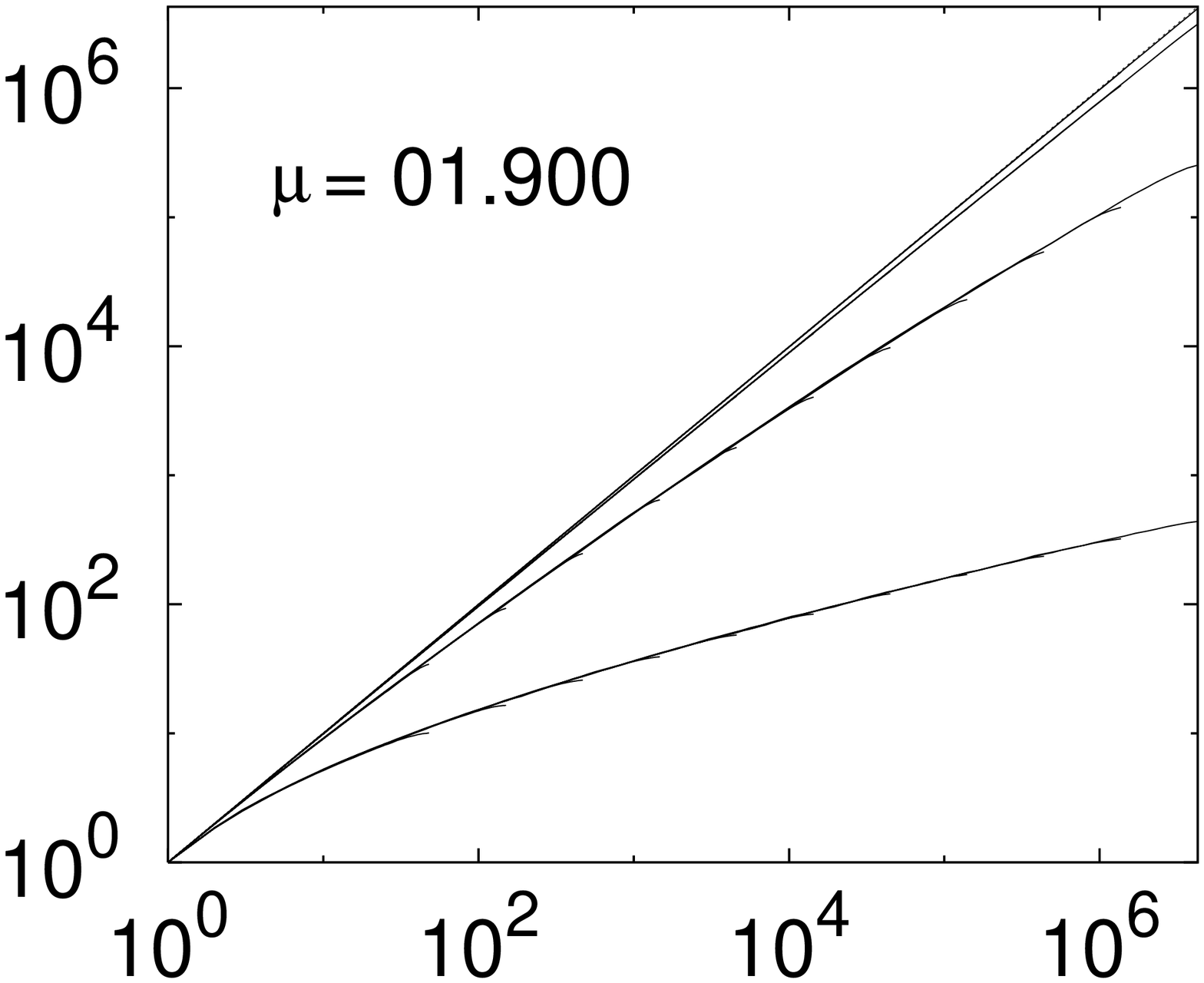,width=5.3cm,angle=270}
}}
\caption{{} Same as \protect \Fig{fig:ell_0a}, for $1<\xpn<2$.}
\label{fig:ell_0b}
\end{figure}
\begin{figure}
\vbox{
\leftline{\hskip -1cm
\psfig{figure=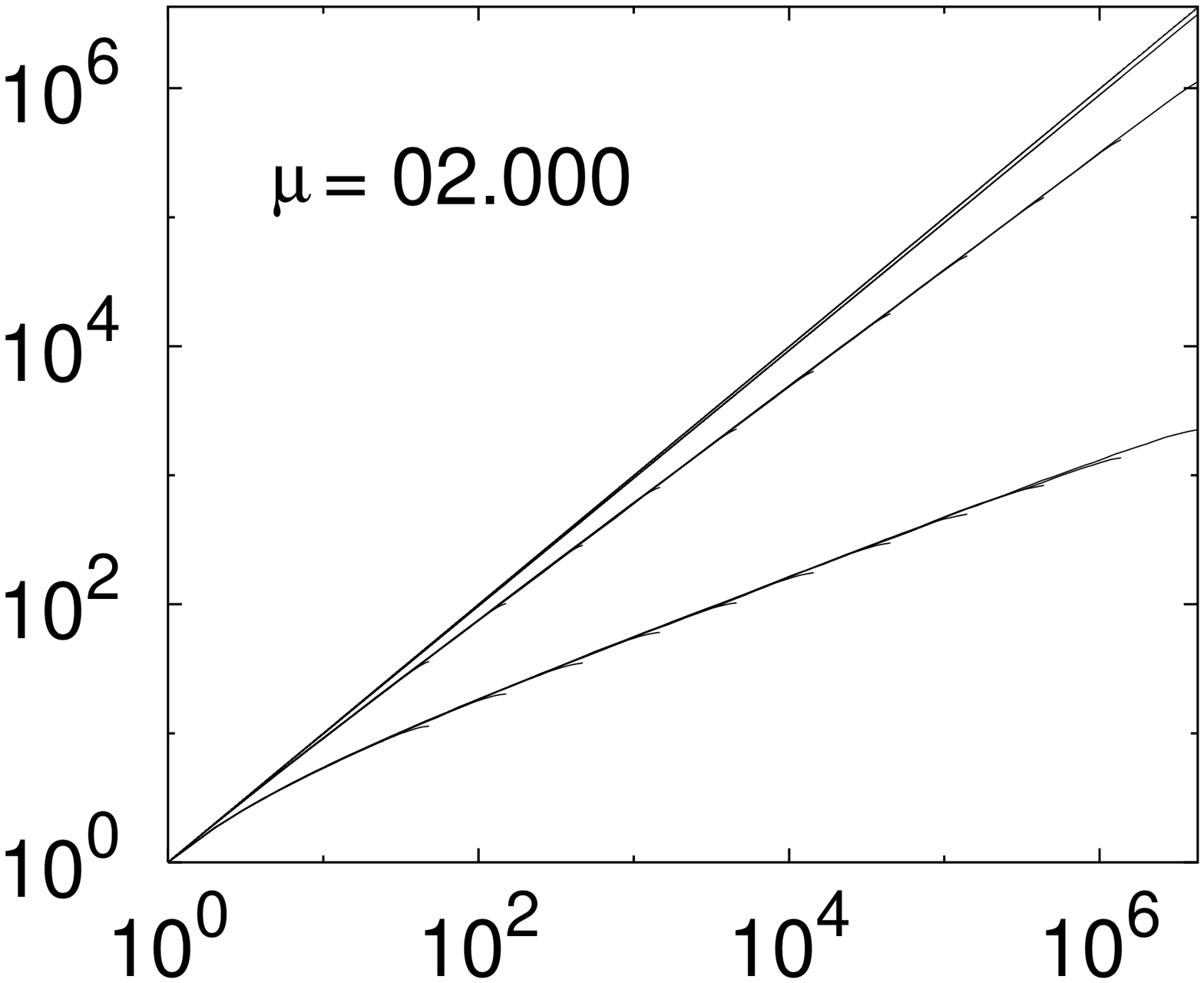,width=5.3cm,angle=270} \hskip -1cm
\psfig{figure=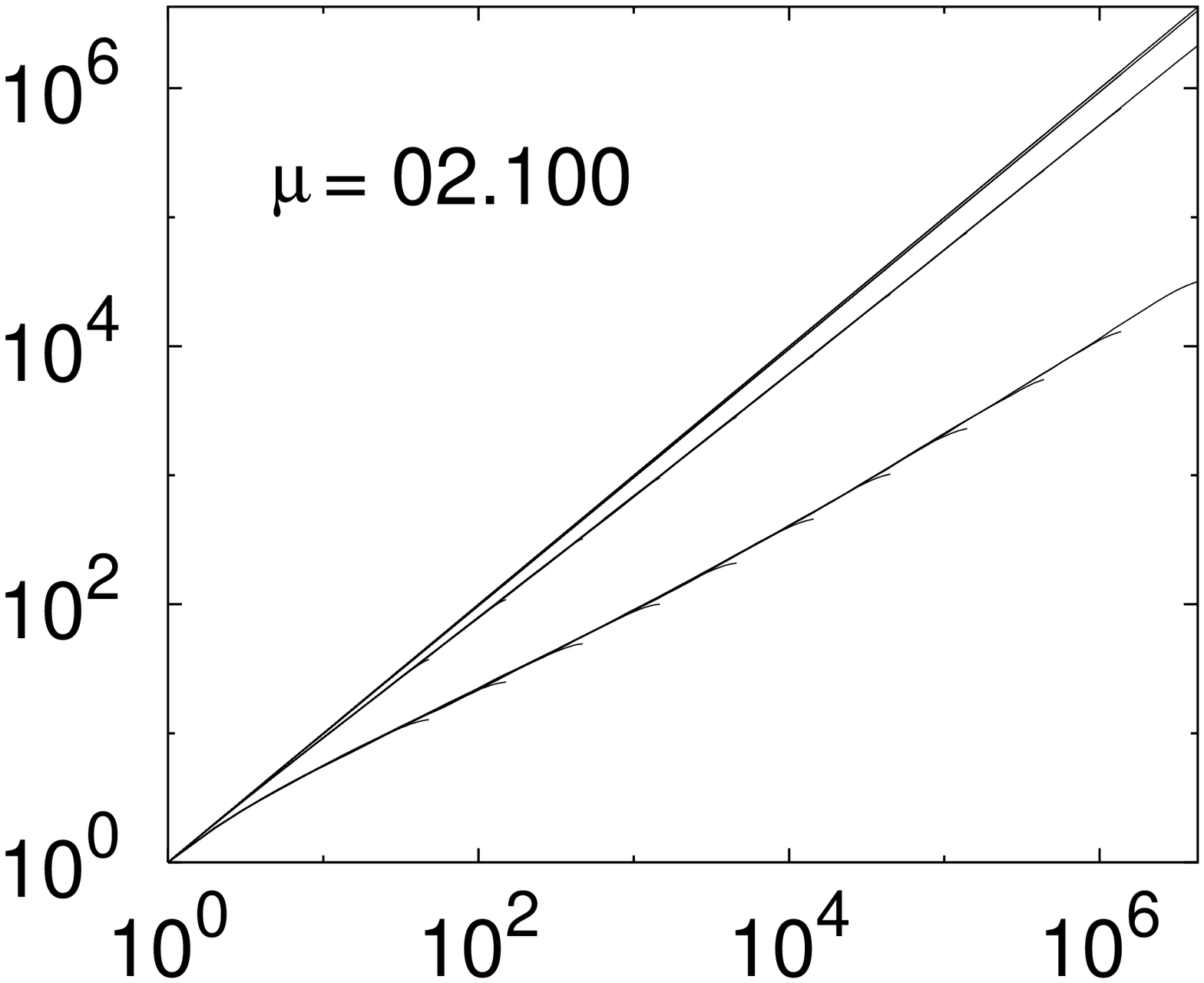,width=5.3cm,angle=270}
}\leftline{\hskip -1cm
\psfig{figure=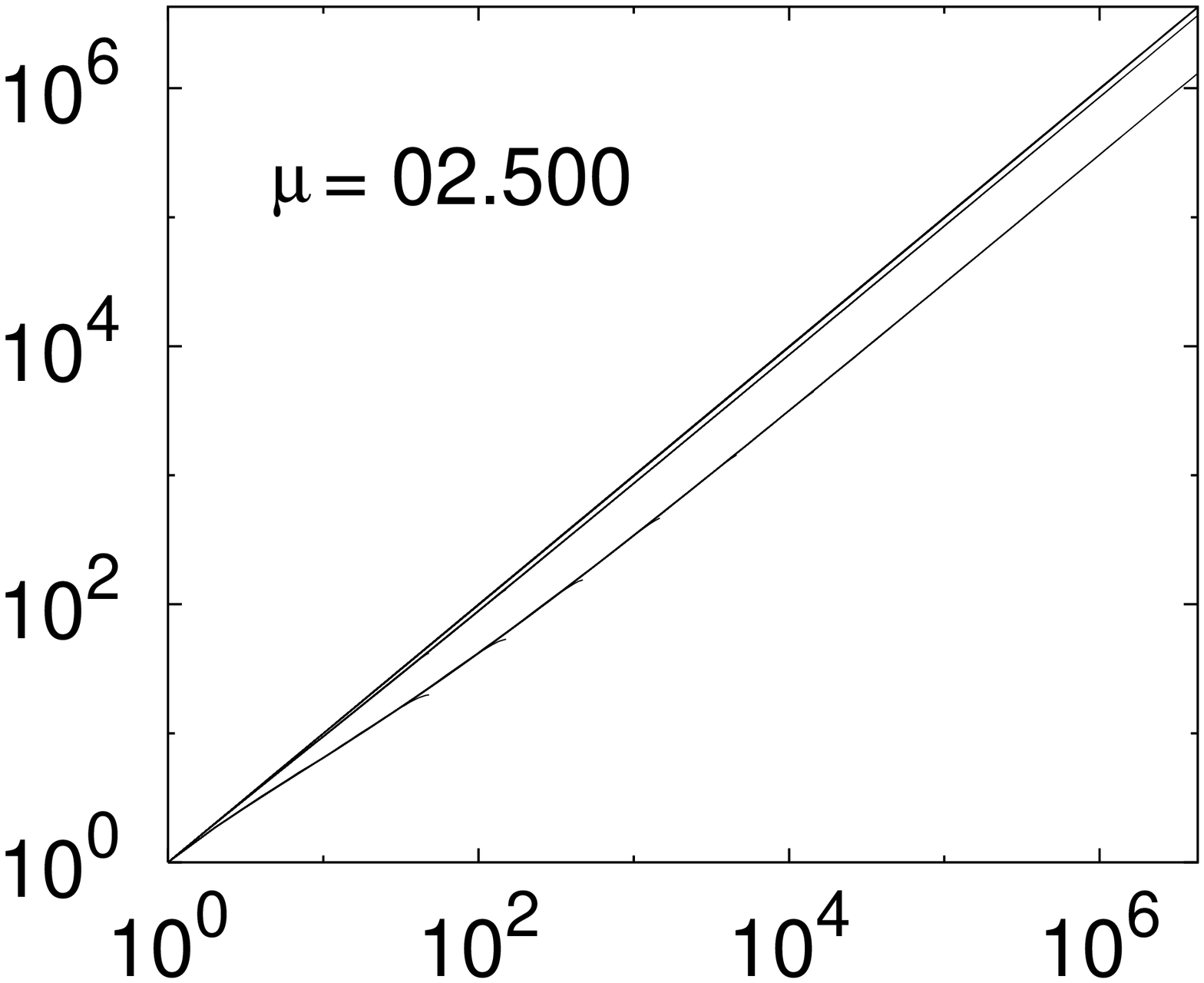,width=5.3cm,angle=270} \hskip -1cm
\psfig{figure=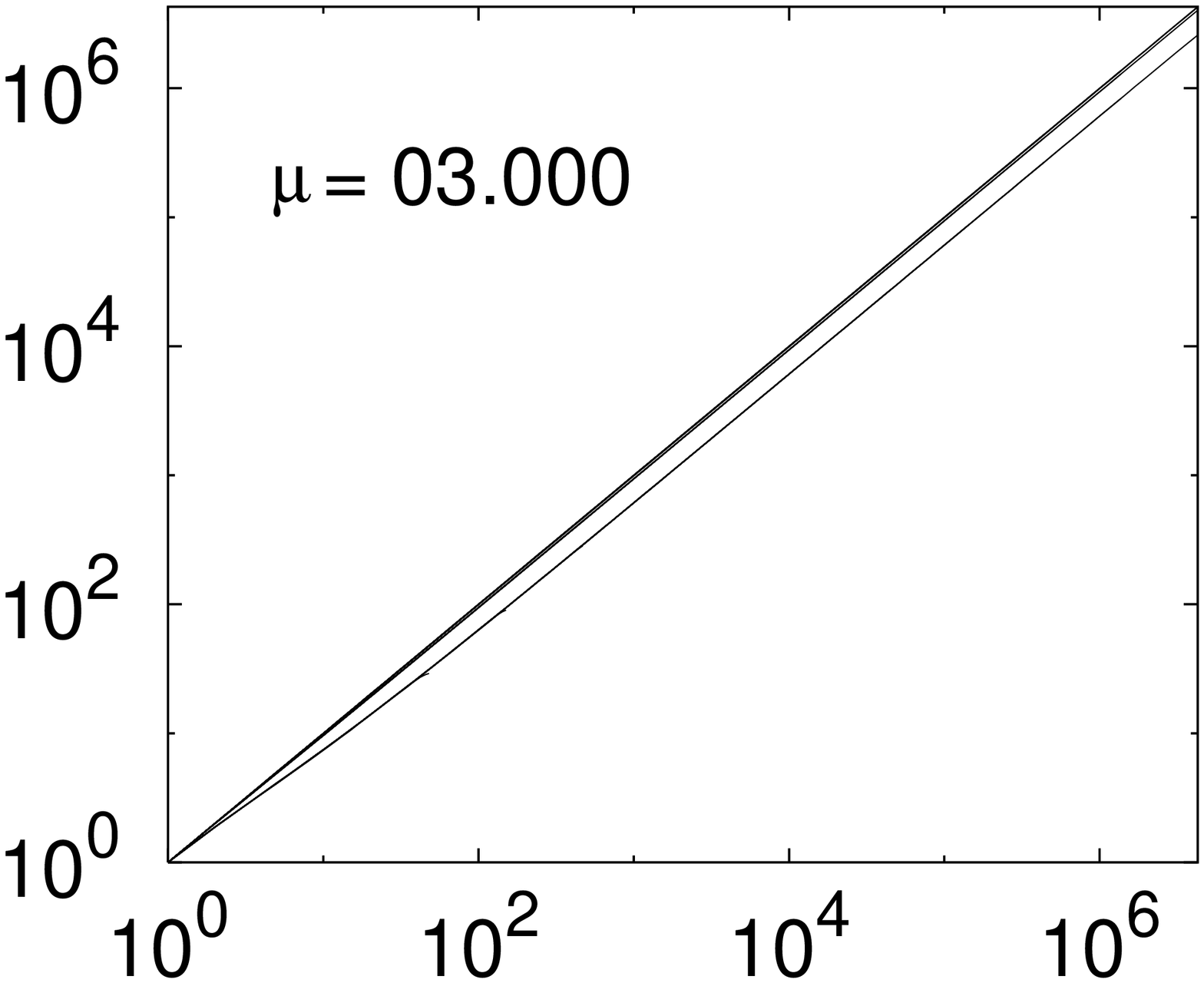,width=5.3cm,angle=270}
}}
\caption{{} Same as \protect \Fig{fig:ell_0a}, for $\xpn \geq2$.}
\label{fig:ell_0c}
\end{figure}
Figures \ref{fig:ell_0a}, \ref{fig:ell_0b} and \ref{fig:ell_0c} show average
shortest-path lengths $\ell(r)$, respectively for the regions: $0\leq \xpn
\leq 1$, $1< \xpn <2$ and $\xpn \geq 2$.
\\
It is apparent in these plots that $\ell(r)$ does not depend on system size
$L$ (only on $p$ and $\xpn$) for $\xpn >1$.  This is consistent with the fact
that the probability $P(r)$ for two sites separated by an Euclidean distance
$r$ to be connected by a LR bond does not depend on $L$ when $\xpn > d$. (See
\Eqn{eq:pofl}). In comparison, when $\xpn <d$ one has that $P(r)$ decays to
zero with system size as $L^{-(d-\xpn)}$.  This scale-dependence in the
connectivity properties is evidenced by the size-dependence of $\ell(r)$ when
$\xpn<1$ in \Fig{fig:ell_0a}.
\\
A second noticeable feature is that for all $\xpn < 2$ a characteristic size
$\xi$ exists with the following property: For $r<<\xi$, $\ell(r) \approx r$,
while for $r>\xi$, $\ell(r)$ grows asymptotically slower than $r$; in general
as $r^{\theta_s}$ with $\theta_s<1$. This characteristic size $\xi$ is a
function of $p$ and $\xpn$ for $1 < \xpn < 2$, but also depends on $L$ for
$\xpn <1$.
\subsection{The $\xpn > 2$ regime}
\label{sec:numericala>2}
As seen in \Sec{sec:model}, for $\xpn >2$ the density of LR bonds rescales to
zero, i.e. $p=0$ is an attractive fixed line. Thus one does not expect LR
bonds to modify the effective geometry of the lattice in this regime. In fact
it is found (\Fig{fig:ell_0c}) that $\ell(r) \propto r$ at large distances,
and thus $d_{eff}=d$ in this regime, although the coefficient of
proportionality depends on $\xpn$ and $p$ in general.  Our directed model
(naive paths) described in \Sec{sec:naive} gives exact results in this regime
as we now show.
\subsubsection{Naive paths when $\xpn>2$}
\begin{figure}[htb]
\leftline{\hskip -0.3cm
\psfig{figure=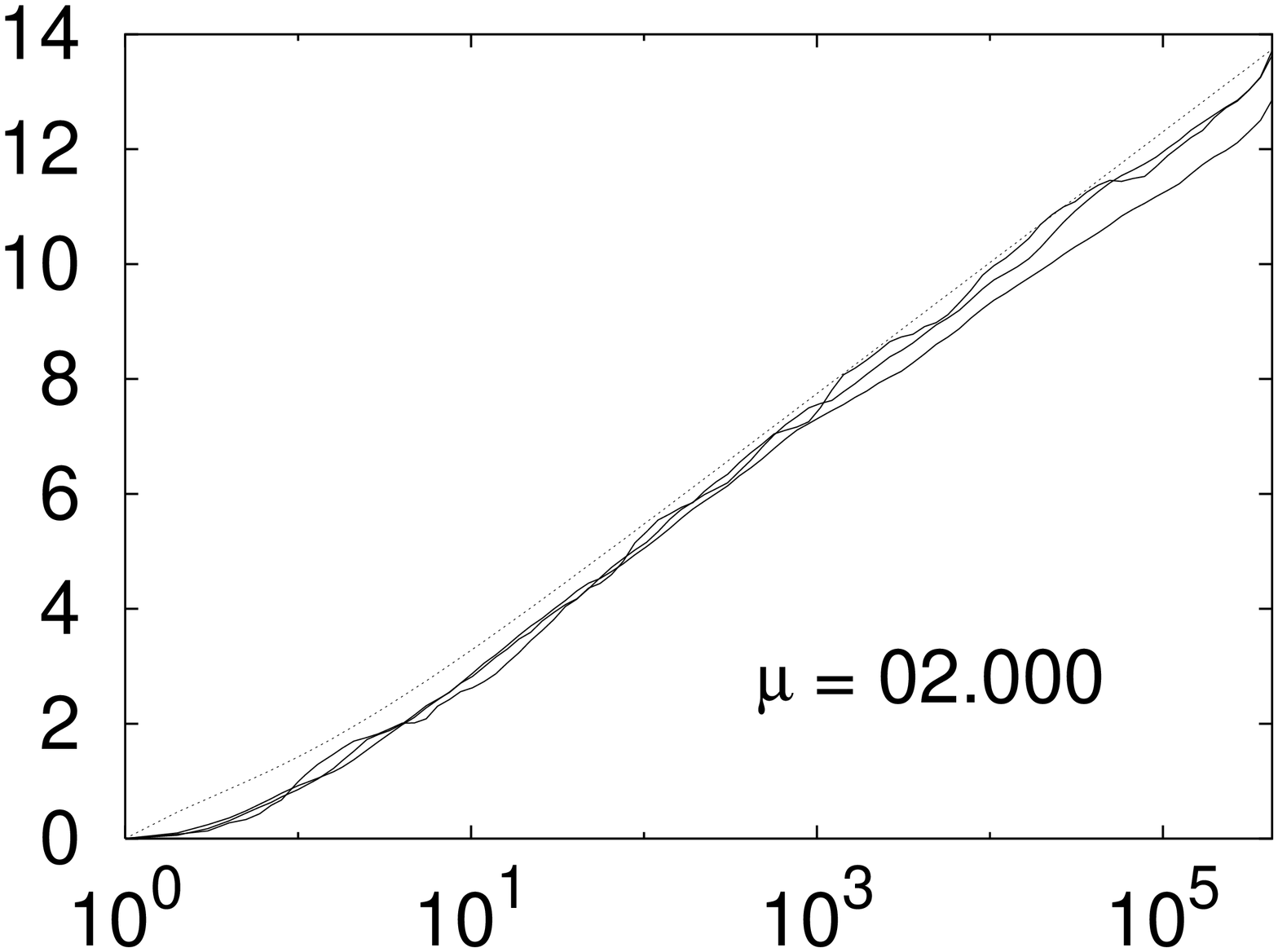,width=5.0cm,angle=270} \hskip -0.5cm
\psfig{figure=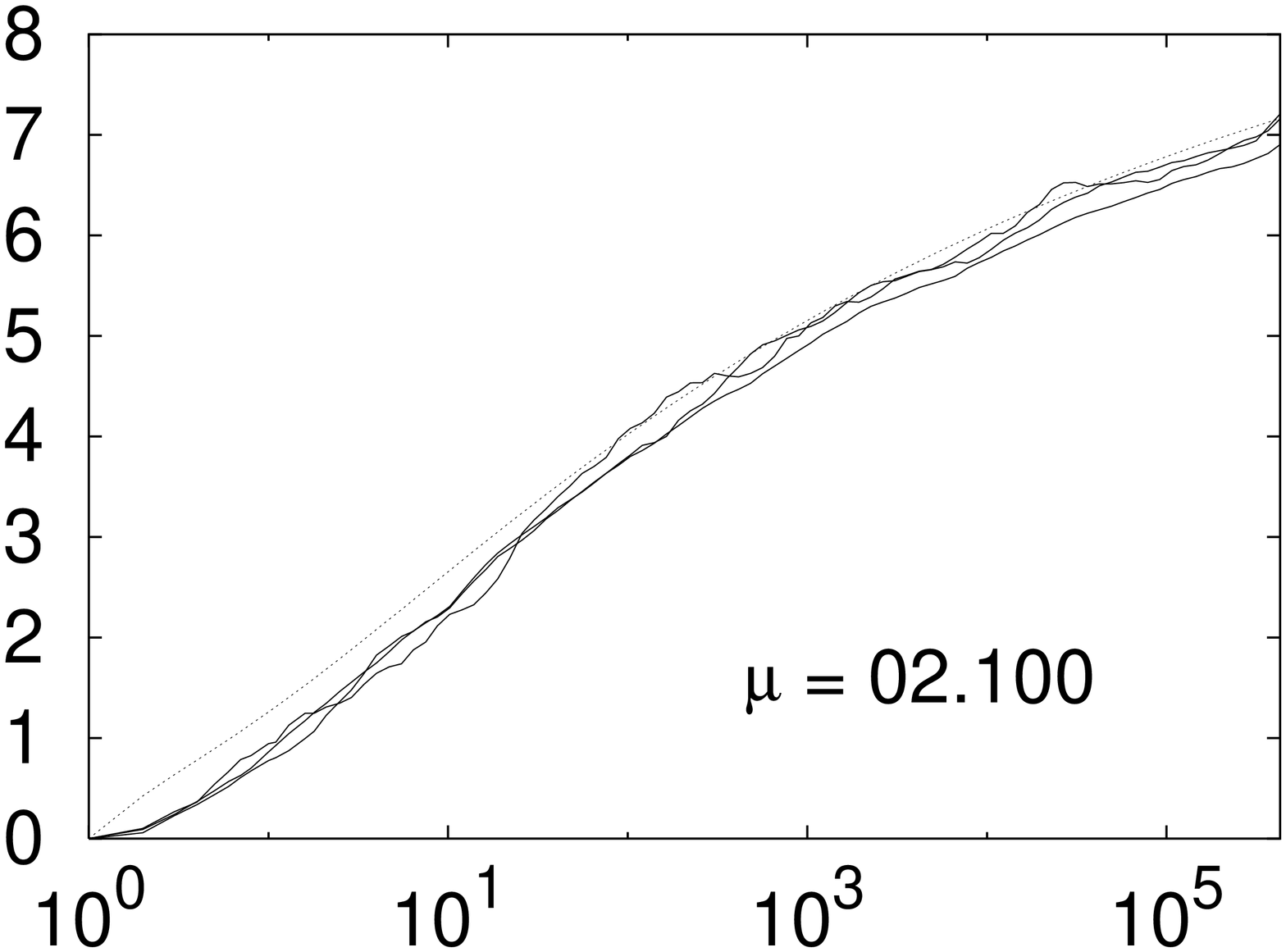,width=5.0cm,angle=270}
}\leftline{\hskip -0.3cm
\psfig{figure=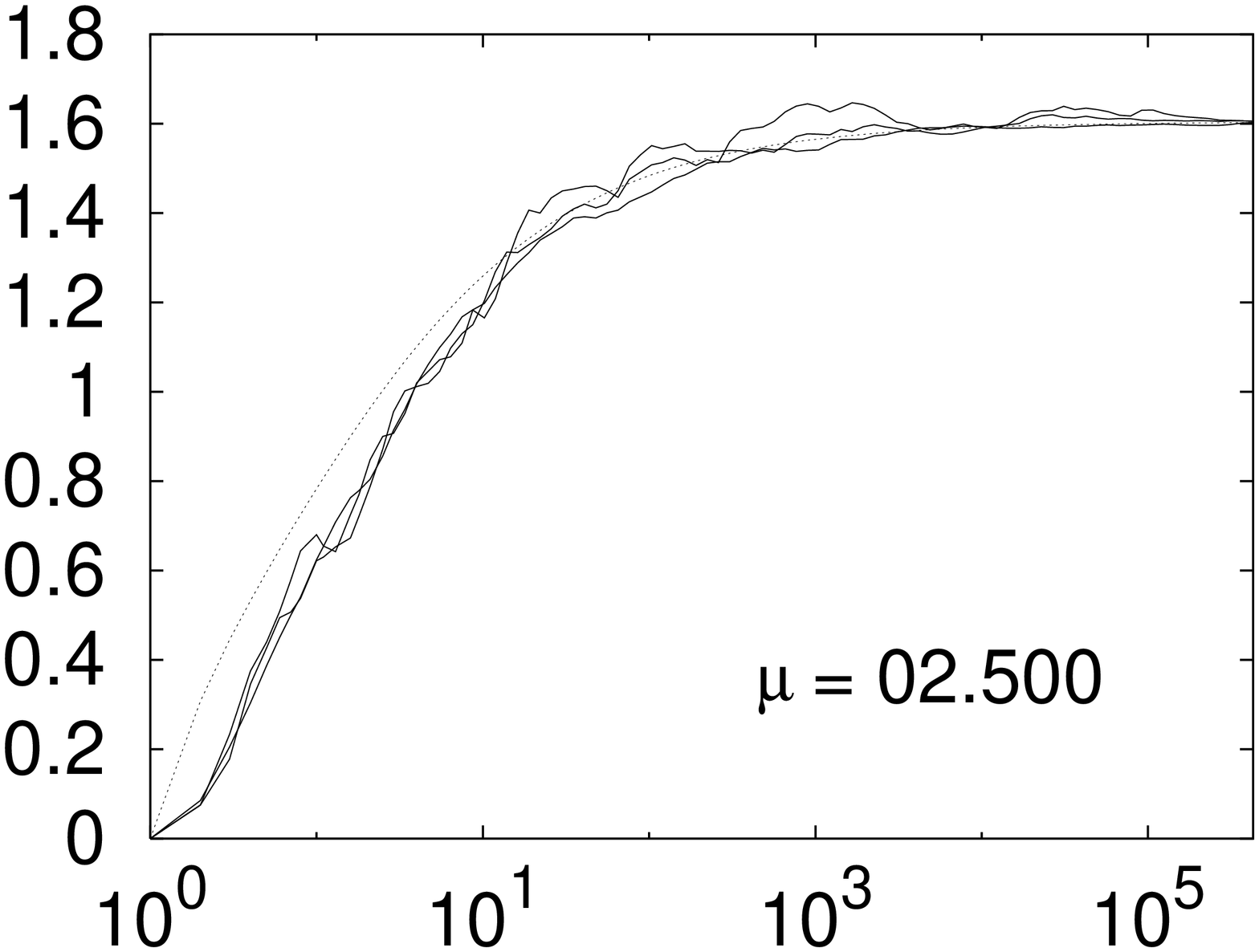,width=5.0cm,angle=270} \hskip -0.5cm
\psfig{figure=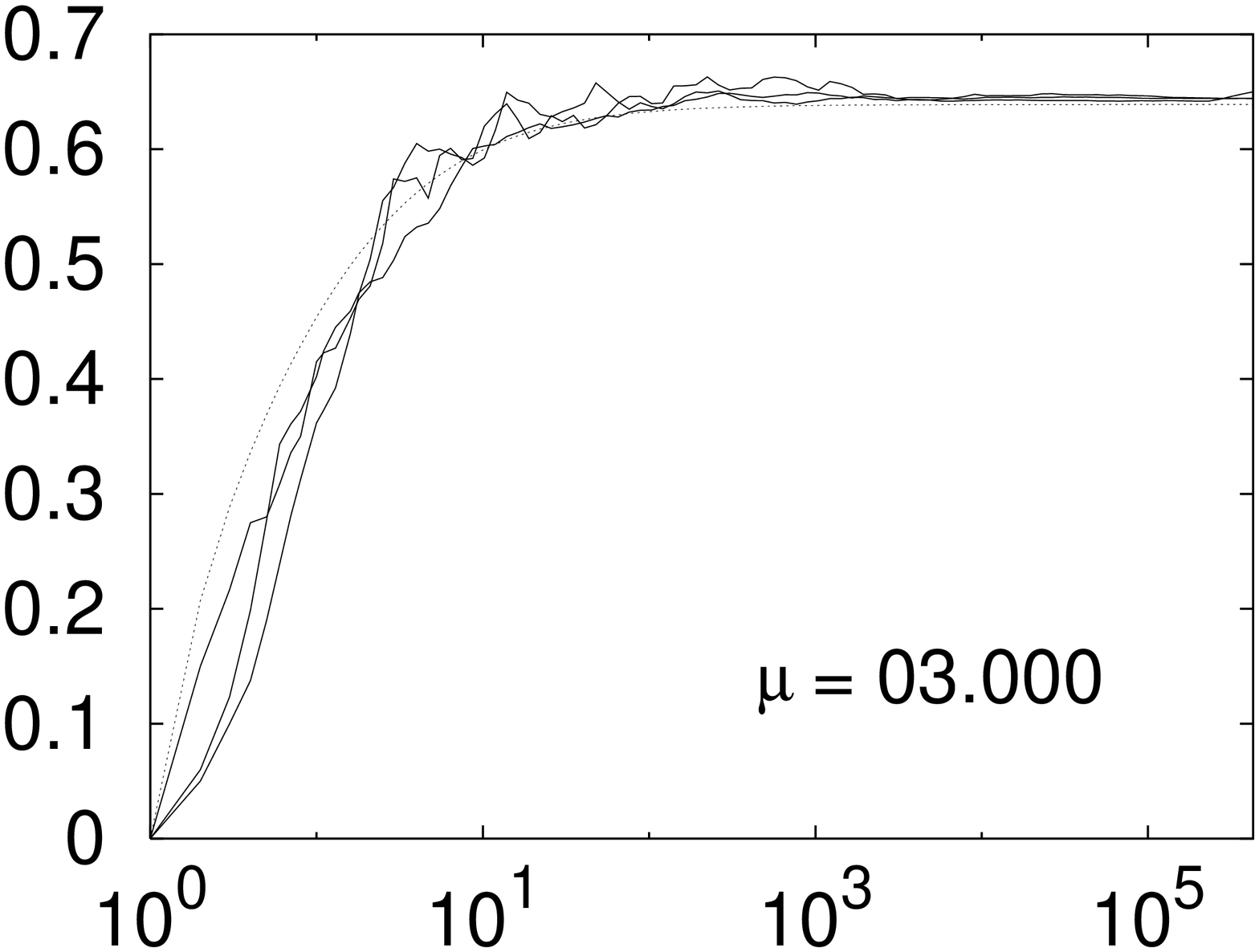,width=5.0cm,angle=270}
}
\caption{{}For $\xpn>2$ shortest path lengths $\ell(r)$ are well approximated
  by \Eqn{eq:solution} with $\Phi$ given by \Eqn{eq:phi}. Shown in this plot
  are our numerical results (solid lines) for $\Phi(r)=p^{-1}(1-\ell(r)/r)$
  for small densities of LR bonds: $p=10^{-3}, 3\times 10^{-2}$ and $10^{-2}$.
  The dashed line indicates our analytic result, \Eqn{eq:phi_approx}. }
\label{fig:ellofr_3}
\end{figure}

When $\xpn >2$, $G(x)$ in \Eqn{eq:naive_formal} grows monotonically from
$G(1)=0$ to $G(\infty)=\bar l -1$. Thus asymptotically $\ell_n(r) =
r/(1+p(\bar l-1))$. In order to obtain the short-distance behavior we may
approximate, to first order in $p(\bar l-1)$,
\begin{equation}
\left [1 + p G(x)\right ]^{-1}  
\approx   1 - p G(x).
\end{equation}
\Eqn{eq:naive_formal} now reads
\begin{equation}
\ell_n(r) \approx r \left [ 1- p \Phi(r) \right ],
\label{eq:solution}
\end{equation}
where 
\begin{equation}
\Phi(r)= \frac{1}{r} \int_1^r G(x) dx
\label{eq:phi}
\end{equation}
is a $p$-independent function which converges to $\bar l-1$ for large $r$.
Equation \ref{eq:phi} can be integrated (See \Sec{apd:naive}), and the
comparison between analytical and numerical results is done in
\Fig{fig:ellofr_3}. The coincidence betweeen the naive-path model and
numerical results is very good even at $\xpn=2$. Thus we conclude that in the
$\xpn \geq 2$ regime and when $p$ is small, shortest-paths are essentially
naive paths.

\subsection{The $1<\xpn<2$ regime}
\label{sec:numerical1<a<2}
In \Sec{sec:rescaling} we saw that $p=0$ is a repulsive fixed point for all
$\xpn <2$ in one dimension.  Because of the rescaling law \Eqn{eq:p-exponent},
one expects a lengthscale $\xi \sim p^{-1/y_p} = p^{-1/(2-\xpn)}$ to be
relevant for the behavior of $\ell(r)$ as $p \to 0$. For $r<<\xi$, the $p=0$
fixed point is dominant (for which $\ell(r)=r$) while for $r>>\xi$ the effects
of LR bonds may become visible ($\ell(r)$ shorter than $r$). 
\subsubsection{Naive paths when $1 < \xpn < 2$}
\label{sec:naive_b}
For $\xpn<2$, $\bar l$ is not well defined. However the average length $\tilde
l_t$ of a LR bond not larger than $r-x_t$ is well defined and given by
\Eqn{eq:alen0}. Notice that $G(r)$ now grows as $r^{2-\xpn}$.
\Eqn{eq:naive_formal} is still valid for naive paths, and one gets in the
limit of large $r$ that $\ell(r) \sim r^{\xpn-1}$, i.e.
$\theta_s^{naive}=\xpn-1$.  It turns out that actual shortest-paths are
shorter than naive paths for $\xpn<2$, i.e $\theta_s^{naive}=\xpn-1$ is only
an upper bound for $\theta_s$ (see \Sec{sec:theta_s}).

Although the naive-path model fails to predict the asymptotic behavior of
$\ell(r)$, it can nevertheless still help us determine the characteristic
length $\xi$ beyond which $\ell(r)/r \to 0$. Keeping just the fastest-growing
term in $G(y)$ (\Eqn{eq:solution}) and equating $pG(\xi) \approx 1$, one gets
$\xi \sim p^{-1/(2-\xpn)}$, in full accordance with rescaling arguments in
\Sec{sec:rescaling} and at the beginning of this section. We show next that
this is verified numerically.
\subsubsection{A single characteristic length $\xi$}
\label{sec:xi}
In this subsection we test the hypothesis that a single lengthscale $\xi(p)$
dictates the behavior of $\ell(r)$ in the limit of small $p$, and show that
for $1\leq \xpn \leq 2$ this lengthscale is $\xi=p^{-1/(2-\xpn)}$, in
accordance with rescaling arguments (\Eqn{eq:p-exponent}) and naive-path
predictions. We propose that, for $p \to 0$,
\begin{equation}
\ell(r,\xpn,p)/\xi = f(\xpn,r/\xi),
\label{eq:scaling}
\end{equation}
where $\xi \sim p^{-\nu}$, and 
\begin{equation}
f(\xpn,x) \propto \left \{
\begin{array}{lrr}
x & \hbox{for} & x<<1 \\ \\
x^{\theta_s(\xpn)} \quad &\hbox{for} & x>>1
\end{array}
\right .
\label{eq:scalingfunc}
\end{equation}
This means that all $p$-dependence of $\ell(r)$ is contained in $\xi(p)$.

\begin{figure}[htb]
\centerline{
\psfig{figure=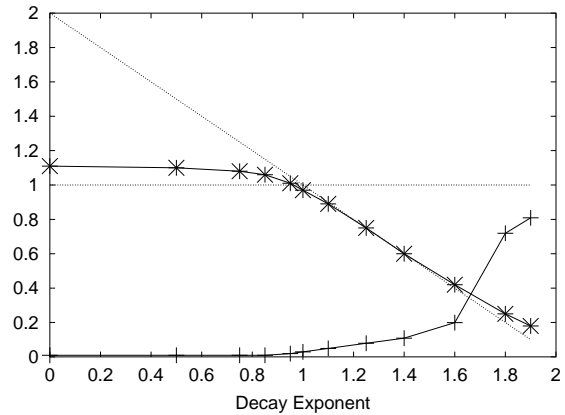,width=8.0cm,angle=270}
}
\caption{{} Numerical estimates for $1/\nu$ (asterisks) and $\theta$
  (plusses), obtained by fitting \Eqn{eq:scalef} to our data for $L=10^7$ and
  $p=0.001, 0.003$ and $0.010$.  The dotted lines are $1/\nu=2-\xpn$ and
  $1/\nu=1$.}
\label{fig:scalefit}
\end{figure}

By comparison with our numerical results we find that $f(x)$ can be well
approximated by $f(x)=x/[1+Cx^{(1-\theta_s)}]$. Therefore
\begin{equation}
\ell(r)/\xi \approx \frac{r/\xi}{1+C \quad (r/\xi)^{1-\theta_s}},
\label{eq:scalef}
\end{equation}
or, equivalently
\begin{equation}
\frac{r}{\ell(r)}-1 \approx C  [rp^\nu]^{1-\theta_s},
\label{eq:fitform}
\end{equation}
provide a good approximation to our numerical results.  We fit
\Eqn{eq:fitform} to our numerical data for $L=10^7$ and $p=0.001, 0.003,
0.010$ simultaneously (using $\nu(\xpn), \theta_s(\xpn)$ and $C(\xpn)$ as
fitting parameters), and find $\nu$ and $\theta_s$ as shown in
\Fig{fig:scalefit}.  These results are entirely consistent with
$1/\nu=(2-\xpn)$ for $1<\xpn<2$.  Larger values of $p$ are found not to follow
\Eqn{eq:scaling} satisfactorily, therefore we must regard this scaling
expression as only valid in the $p\to 0$ limit.

\begin{figure}[hbt]
\vbox{
\leftline{\hskip -1cm
\psfig{figure=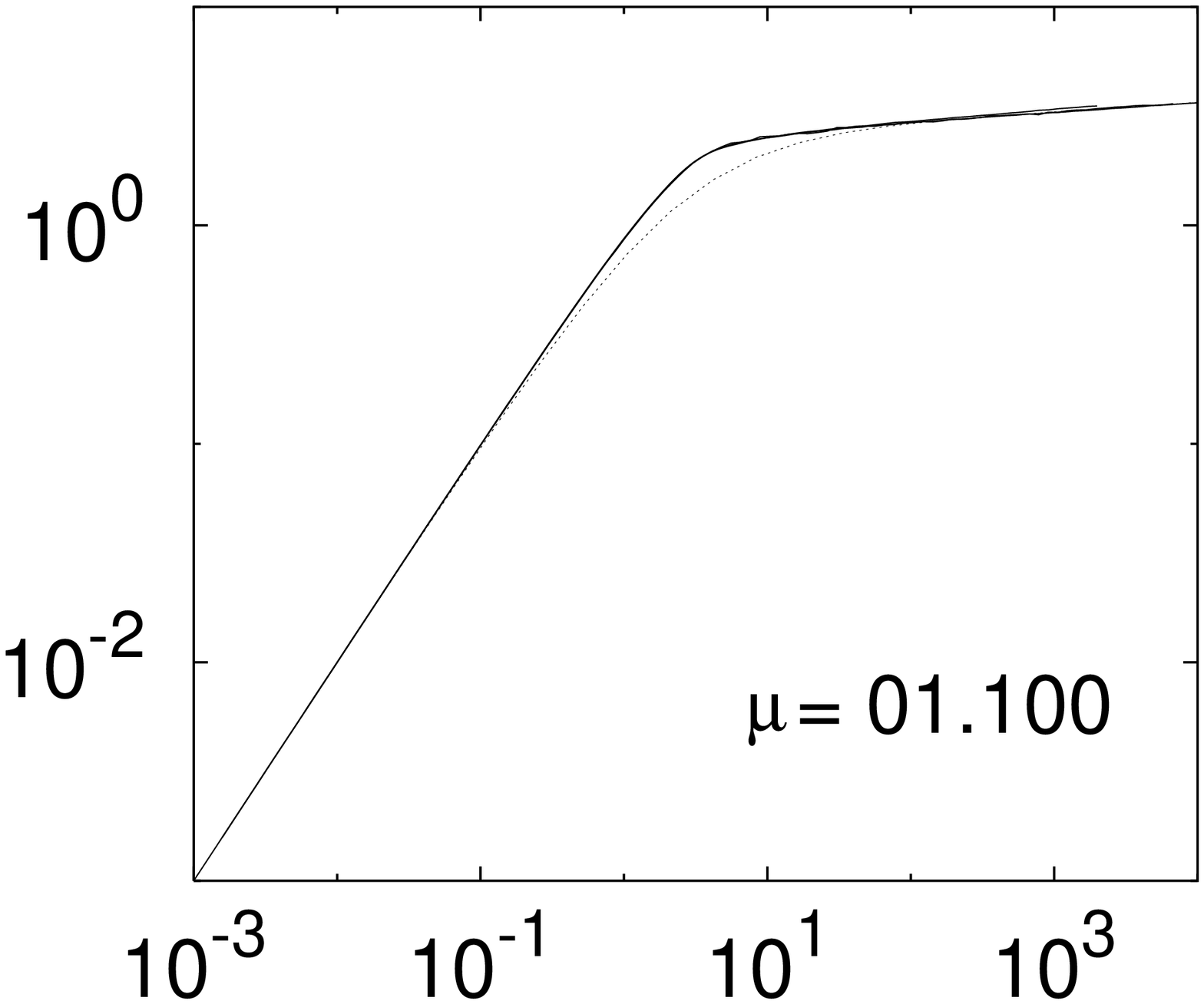,width=5.3cm,angle=270} \hskip -1cm
\psfig{figure=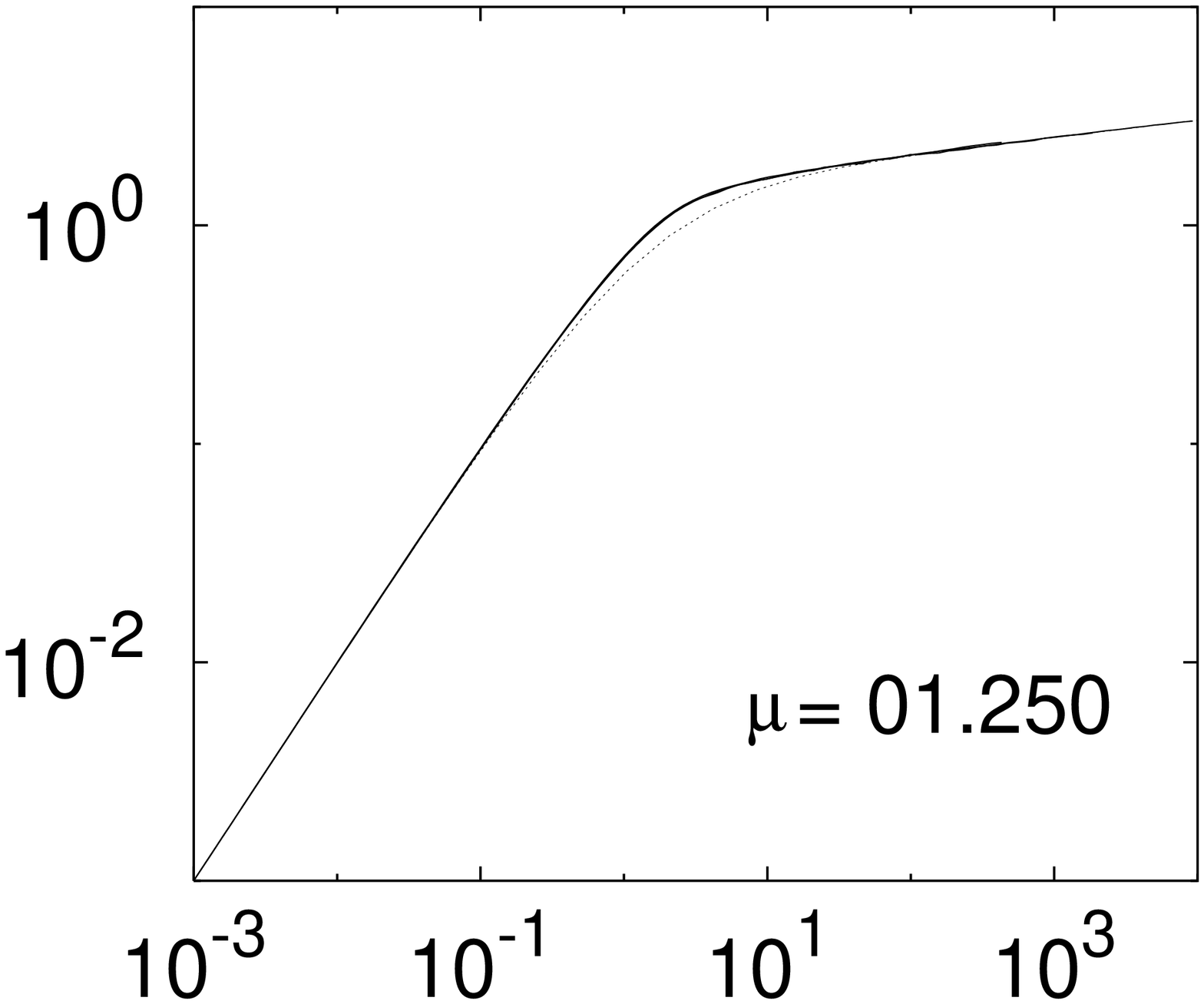,width=5.3cm,angle=270}
}\leftline{\hskip -1cm
\psfig{figure=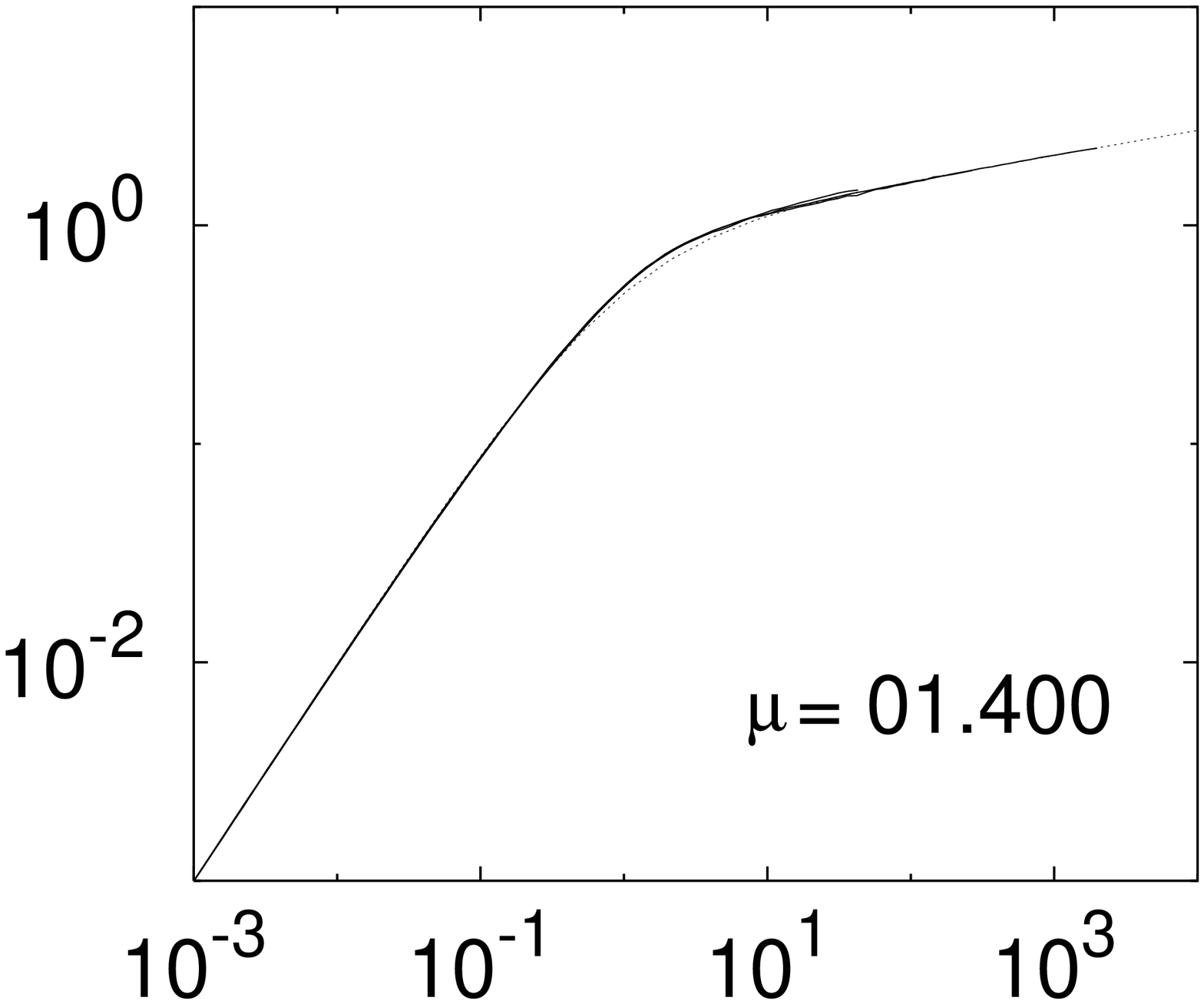,width=5.3cm,angle=270} \hskip -1cm
\psfig{figure=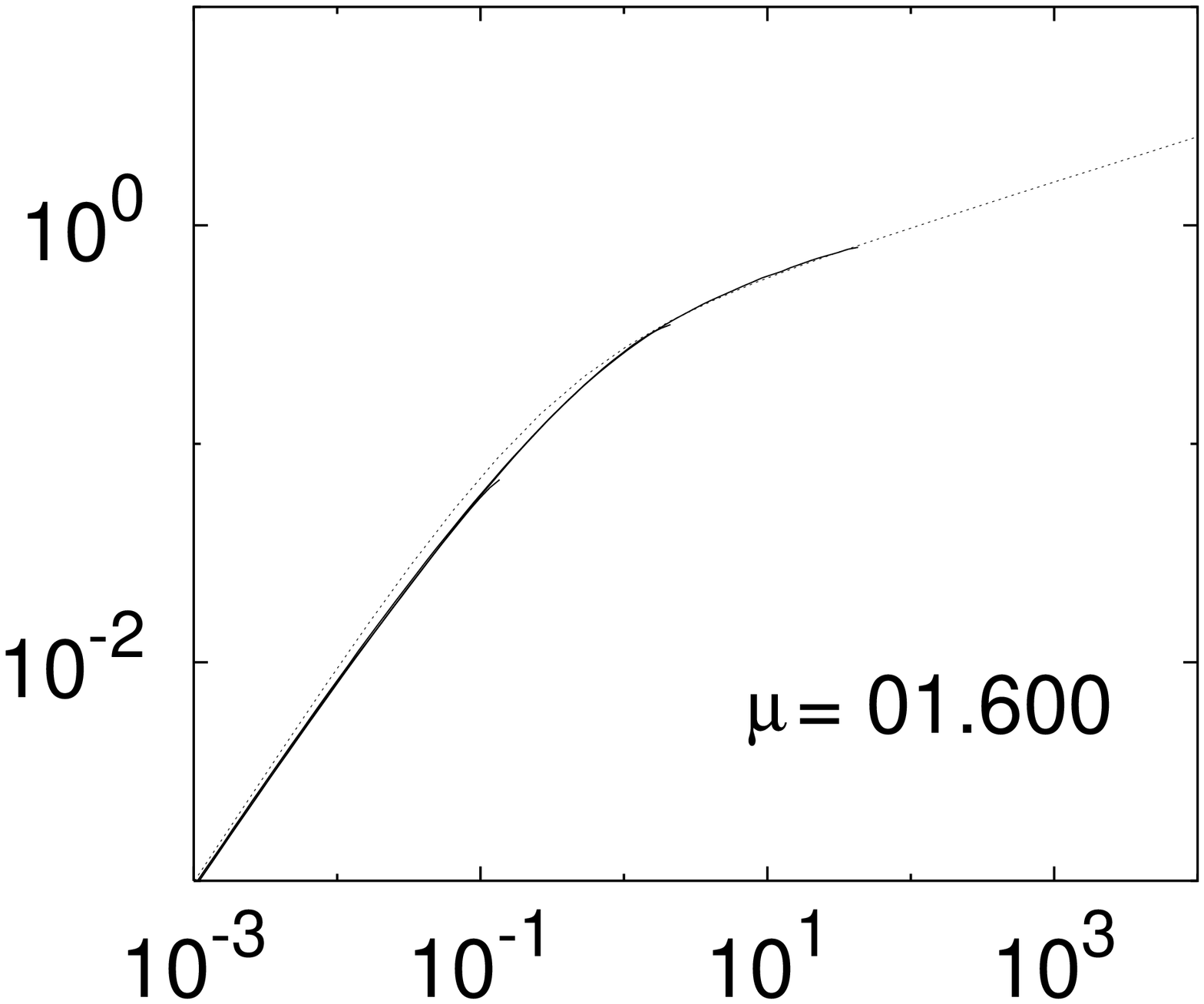,width=5.3cm,angle=270}
}}
\caption{{} Data collapse of $\ell(r)$, showing plots of $\ell/\xi$ vs $r/\xi$
  with $\xi(p)=p^{-1/(2-\xpn)}$, for $p=0.001, 0.003$ and $0.010$. The dashed
  line is our approximate expression \Eqn{eq:scalef}. For larger values of
  $\xpn<2$ and the same values of $p$, the characteristic size is much larger
  than $10^7$.}
\label{fig:ellofr_1}
\end{figure}

A plot of $\ell(r)/\xi(p)$ vs.  $r/\xi$ is shown in \Fig{fig:ellofr_1} for
$p=0.001, 0.003$ and $0.010$. The fact that all three values of $p$ collapse
neatly onto one single curve suffices to verify the correctness of our scaling
ansatz \Eqn{eq:scaling} for small $p$. The specific form of $f(x)$ chosen in
\Eqn{eq:scalef} should however only be regarded as empiric.

Although for $\xpn<1$ we do not expect \Eqn{eq:fitform} to hold (since then
$\xi$ has an additional $L$-dependence not included in these expressions, see
\Sec{sec:numerical0<a<1}), a fit of the data gives $\nu \approx 1$, indicating
that the $p$-dependence of the characteristic size $\xi$ is of the form
$p^{-1}$ in this region. This is again consistent with \Eqn{eq:pscalinglaw}.
We will discuss the regime $\xpn<1$ in detail later in
\Sec{sec:numerical0<a<1}.
When $p$ is small and $\xpn$ is close to $2$, $\xi$ grows too large.
Consequently neither $\xi$ nor $\theta$ can be correctly estimated for $\xpn >
1.6$. Consider for example $p=10^{-2}$. One then has $\xi \sim 10^5$ for
$\xpn=1.6$, but $\xi \sim 10^{10}$, well beyond our present reach, for
$\xpn=1.8$. Thus the estimates for $\theta$ and $1/\nu$ in \Fig{fig:scalefit}
are to be disregarded for $\xpn>1.6$.
\subsubsection{Asymptotic exponent $\theta_s$}
\label{sec:theta_s}
\begin{figure}
\leftline{\hskip -0.3cm
\psfig{figure=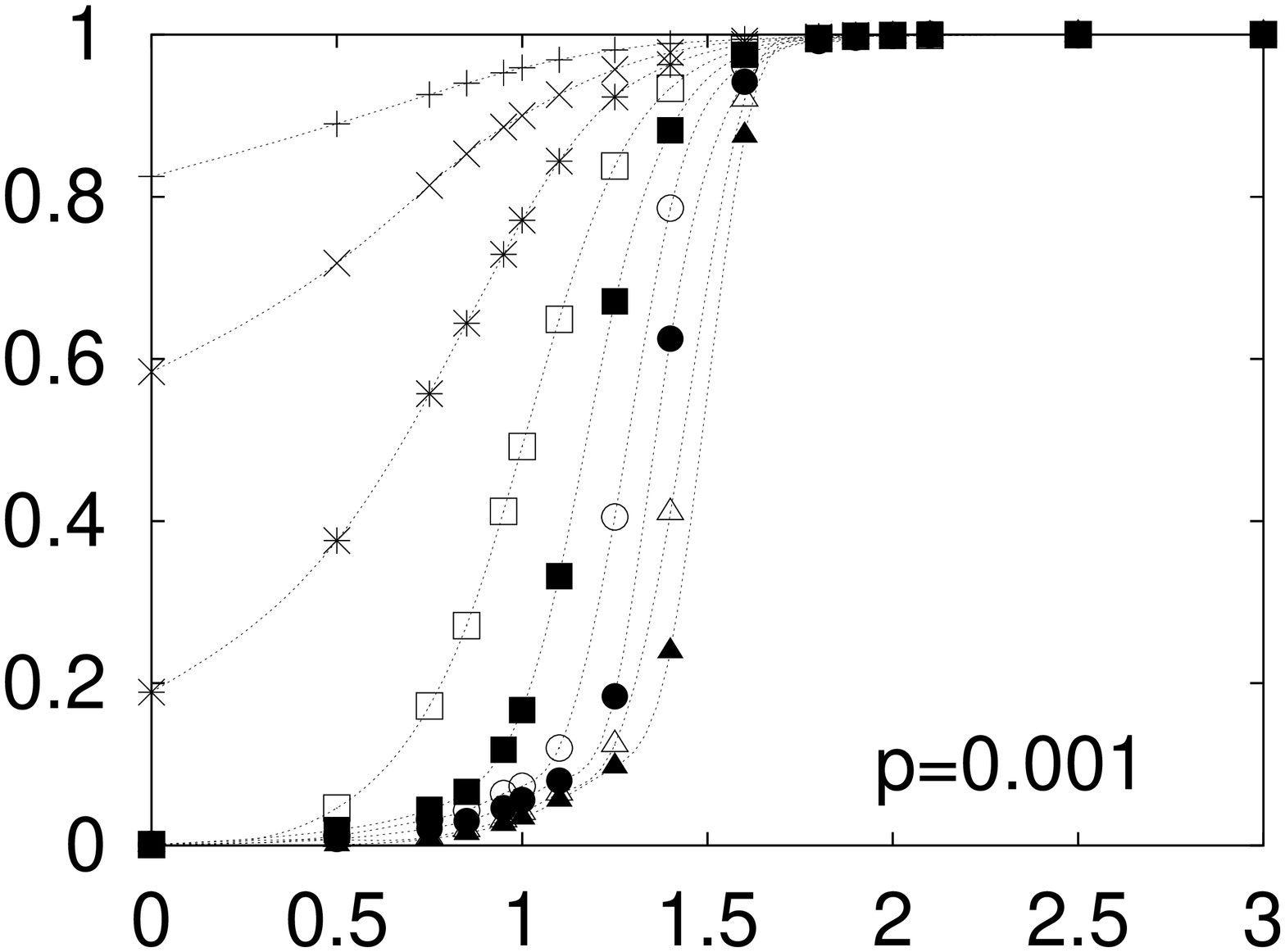,width=5.0cm,angle=270} \hskip -0.5cm
\psfig{figure=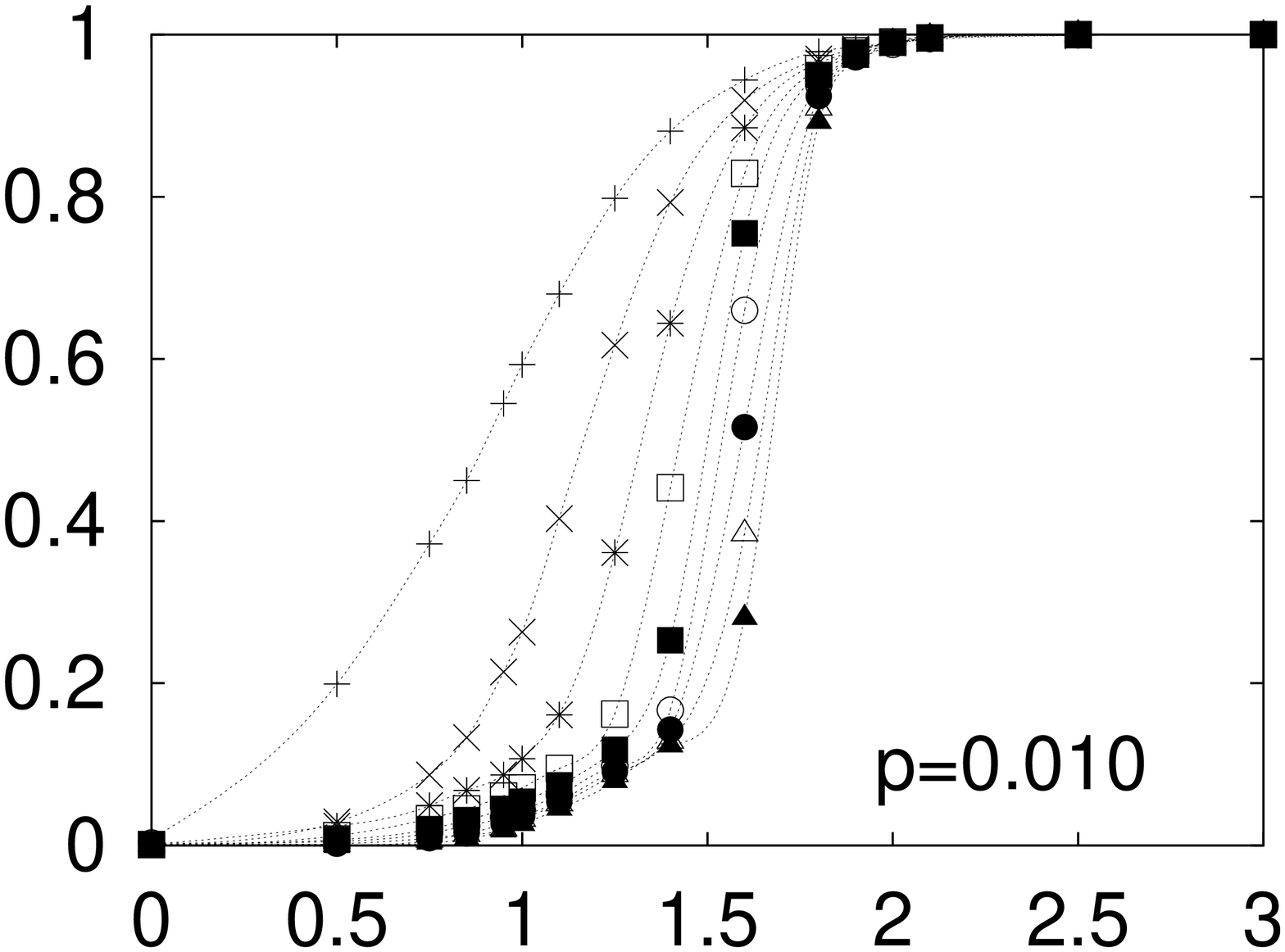,width=5.0cm,angle=270}
}
\leftline{\hskip -0.3cm
\psfig{figure=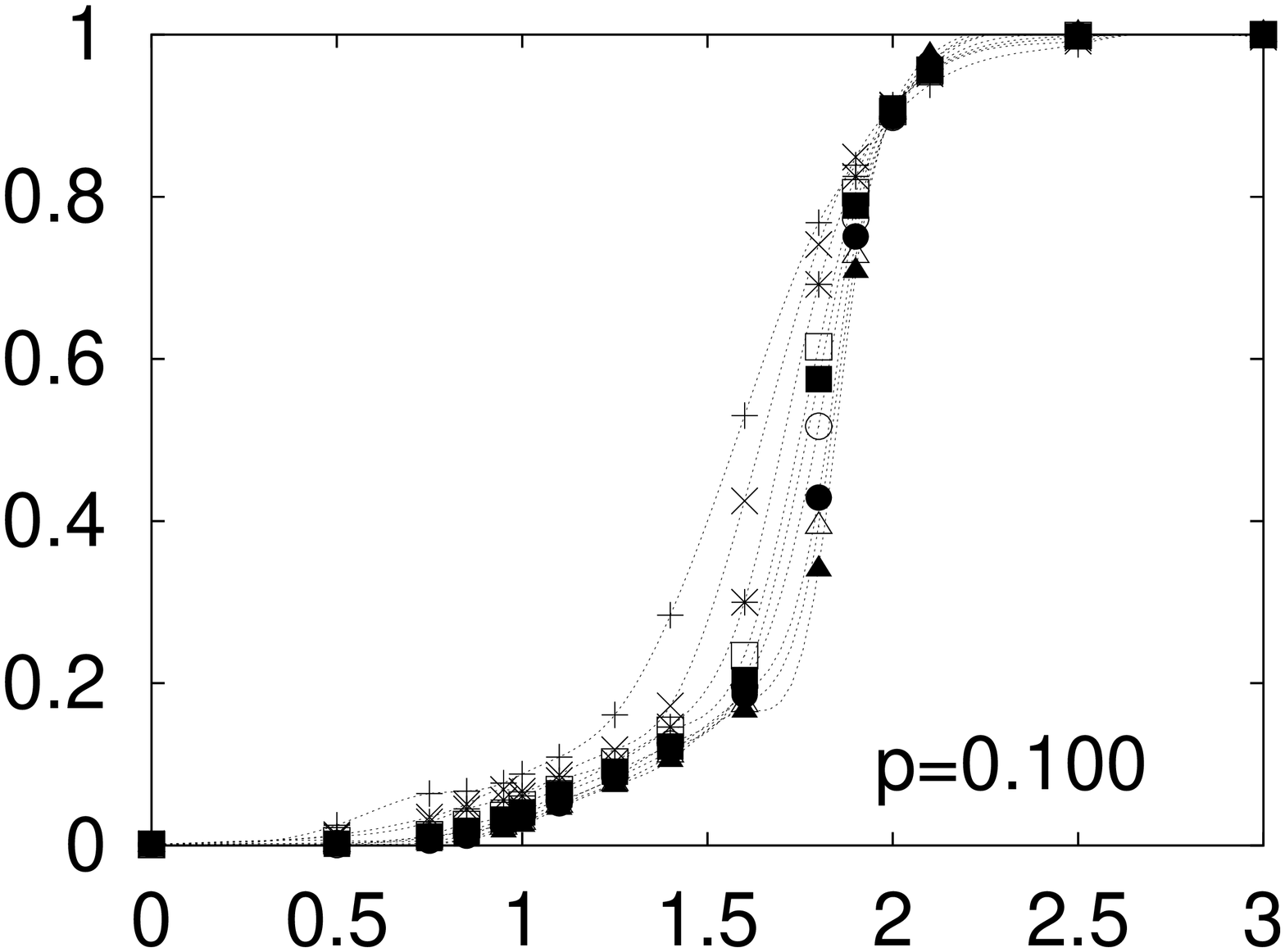,width=5.0cm,angle=270} \hskip -0.5cm
\psfig{figure=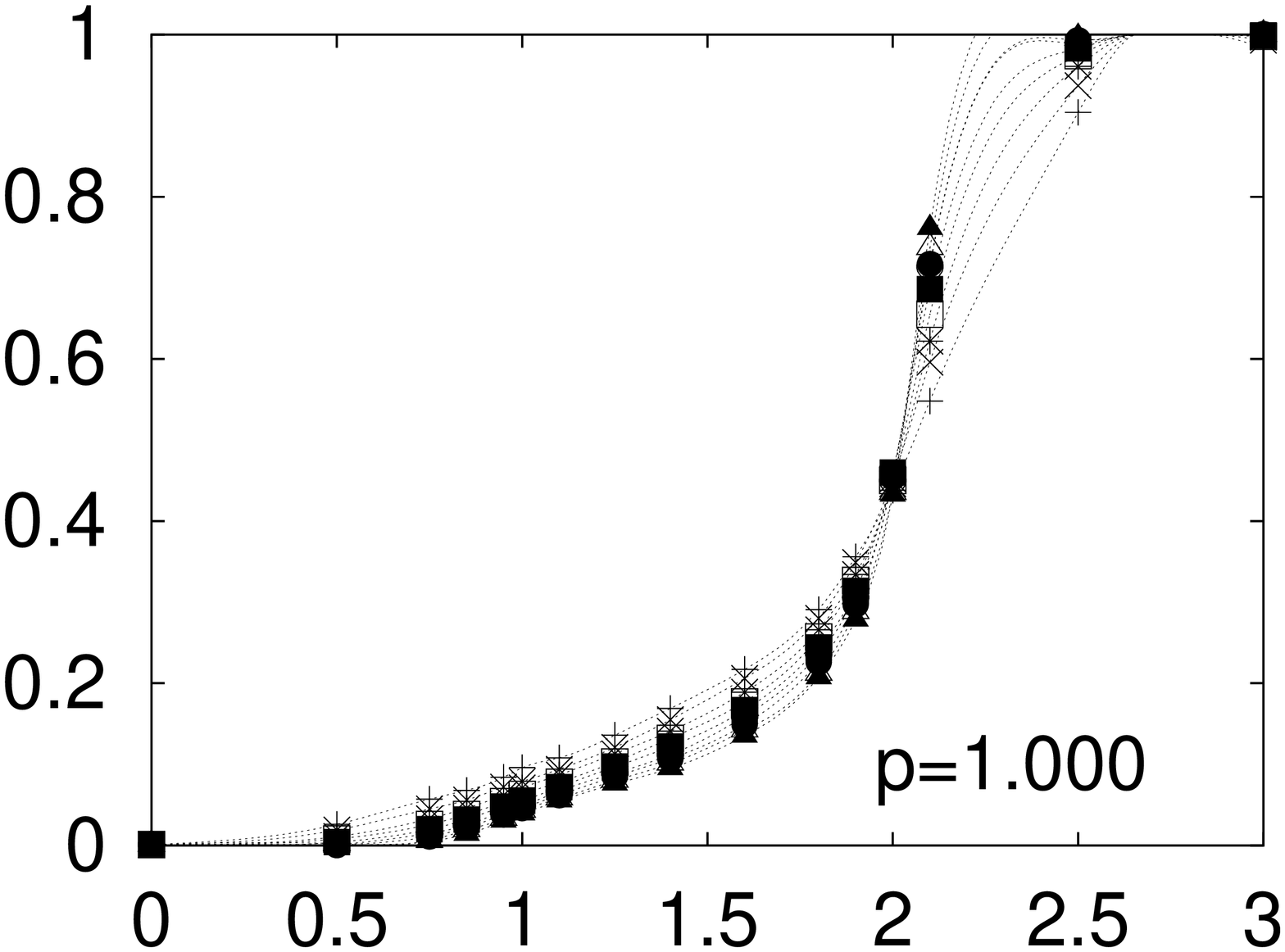,width=5.0cm,angle=270}
}
\caption{{}
  Asymptotic exponent $\theta_s$ obtained from power-law fit of the large-$r$
  behavior of $\ell(r)$, for $L$ of the form $L_k=10^{3+k/2}, k=0,1\ldots,8$ ,
  and for the values of densities of LR bonds $p$ indicated in the respective
  plots. Lines are guides to the eye. }
\label{fig:theta_s1}
\end{figure}

When $r>>\xi$, we find that $\ell(r)$ grows asymptotically as $r^{\theta_s}$.
The shortest-path dimension $\theta_s$ depends on $\xpn$ only, goes to zero as
$\xpn \to 1^{+}$ and jumps discontinuously to $\theta_s=1$ at $\xpn=2^{-}$.
We estimate $\theta_s$ by two different methods. A simple power-law fit of the
large-$r$ behavior of $\ell(r,\xpn,p)$ gives the estimates shown in
\Fig{fig:theta_s1} for $L$ ranging from $10^3$ to $10^7$ and several values of
$p$. Strong finite-size corrections affect the smaller values of $p$, for
which $\xi >> L$ when $\xpn \to 2$. However for large $L$ all these estimates
are seen to converge to similar values within numerical accuracy.

\begin{figure}[htb]
\centerline{
\psfig{figure=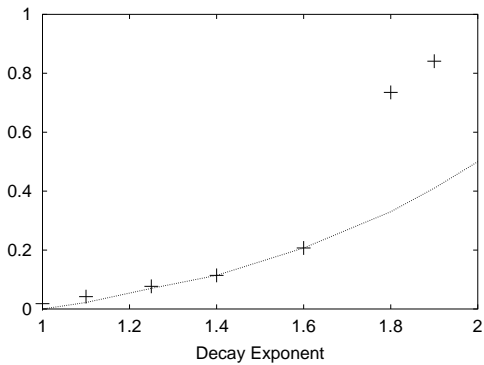,width=7.0cm,angle=270}
}
\caption{{} Numerical estimates for the asymptotic exponent
  $\theta_s$ (plusses) in \protect \Eqn{eq:scalef}, resulting from fits of our
  data for $L=10^7$ and $p=0.001, 0.003$ and $0.010$ with
  $\xi=p^{-1/(2-\xpn)}$. The rightmost two points, for $\xpn$ larger than
  $1.6$, suffer from strong finite-size effects and should be disregarded. The
  dotted line sketches what we believe is the true value of $\theta_s(\xpn)$.
  The discontinuity at $\xpn=2$ is suggested by the behavior of the $p=1.0$
  results in \protect{\Fig{fig:theta_s1}}.  }
\label{fig:scalefit2}
\end{figure}

The second method chosen to estimate $\theta_s$ consists in fitting our
numerical data using \Eqn{eq:fitform} but with $\nu=1/(2-\xpn)$ instead of
taking $\nu$ as a fitting parameter as in \Fig{fig:scalefit}. Fits of our data
for $L=10^7$ and $p=0.001,0.003,0.010$ produce the values of $\theta_s$ shown
in \Fig{fig:scalefit2}. Again the results obtained for $\xpn>1.6$ are to be
disregarded since $\xi$ is much larger than $L$ for these values of $p$.
A naive interpretation of the results in \Fig{fig:theta_s1}, for any fixed
value of $L$, could lead one to believe that the transition between linear
behavior ($\ell(r)\propto r$) and sublinear behavior ($\ell(r)/r \to 0$ for
$r\to \infty$) happens at a $p$-dependent boundary $\xpn_c(p)$~\cite{SCS01}.
However, a more careful numerical analysis shows that this transition happens
at $\xpn_c=2$ for all $p$ in the thermodynamic limit, as predicted by
rescaling arguments (\Sec{sec:rescaling}) and the naive-path model
(\Sec{sec:naive}).
This appears to be in partial disagreement with recent work of Sen and
Chakrabarti (SC)~\cite{SCS01}, where the ``regular lattice behavior''
($\ell(L)\sim L$) is claimed to extend below $\xpn=2$ for small values of $p$.
SC explain what they call the lack of small-world behavior in lattice polimers
as being a consequence of the small number of LR connections (small $p$).
Based on the analysis of $\ell(L)$ on relatively small ($L=10^4$) systems, SC
conclude that there is a $p$-dependent phase boundary $\xpn_c(p)<2$, and show
that several lattice polymer models lay marginally on the regular lattice
($\ell(L) \propto L$) side of this boundary.  Our extensive numerical results
and analytic considerations however show that $\xpn=2$ is the critical decay
rate below which $\ell(r) << r$, for \emph{any} density $p$ of LR bonds.  The
$p$-dependent boundary that SC observe is just a logarithmically slow
finite-size effect. At sufficiently low values of $p$, and for $\xpn$ close to
but lower than two, the characteristic length $\xi(\xpn,p)$ is larger than $L$
and thus $\ell(L) \propto L$. Equating $\xi=p^{-1/(2-\xpn)}=L$, one obtains an
apparent boundary $\xpn^{*}(p)=2-\log(1/p)/\log(L)$, which converges
logarithmically slow to $\xpn_c=2$. Replacing $L=10^4$, this last expression
follows closely the boundary reported by SC in Figure 3 of~\cite{SCS01}.
\\
There is a second aspect of \cite{SCS01} with which our findings seem to be in
disagreement. According to SC, there are only two phases regarding the
asymptotic behavior of $\ell(r)$. A logarithmic phase, $\ell(L) \propto
\log(L)$, for $\xpn<\xpn^{*}(p) \approx 2$, and a linear phase for $\xpn >
\xpn^{*}(p)$. Our numerical evidence however suggests a more complex scenario.
For $1<\xpn<2$ we find that $\ell(r) \propto r^{\theta_s}$ with $\theta_s$
small but nonzero (\Fig{fig:scalefit2}), and only for $\xpn<1$ 
$\ell$ becomes logarithmic (See \Sec{sec:numerical0<a<1}).  
\subsection{The $0 \leq \xpn<1$ regime}
\label{sec:numerical0<a<1}
\begin{figure}
\vbox{
\leftline{\hskip -1cm
\psfig{figure=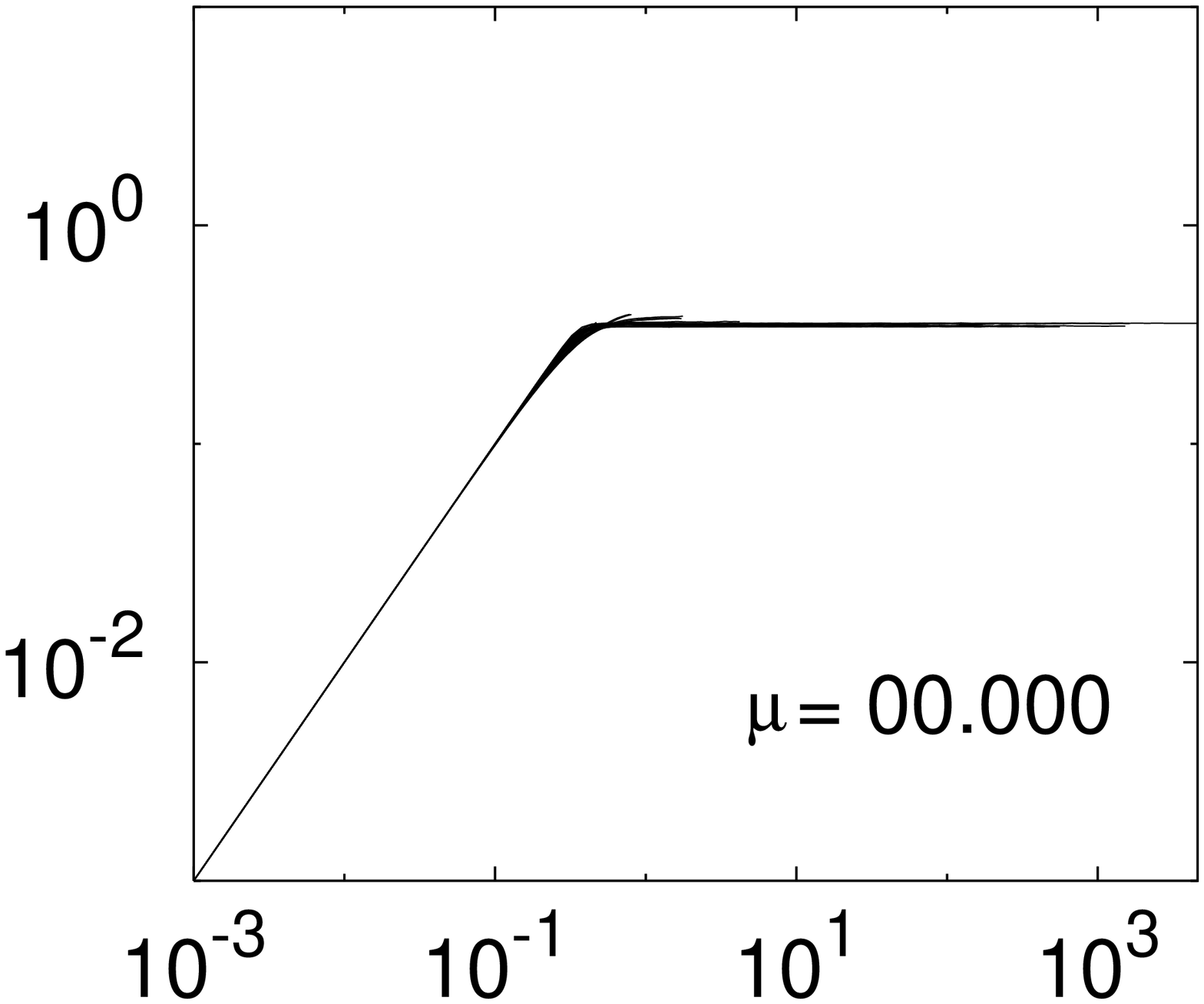,width=5.3cm,angle=270}\hskip -1cm
\psfig{figure=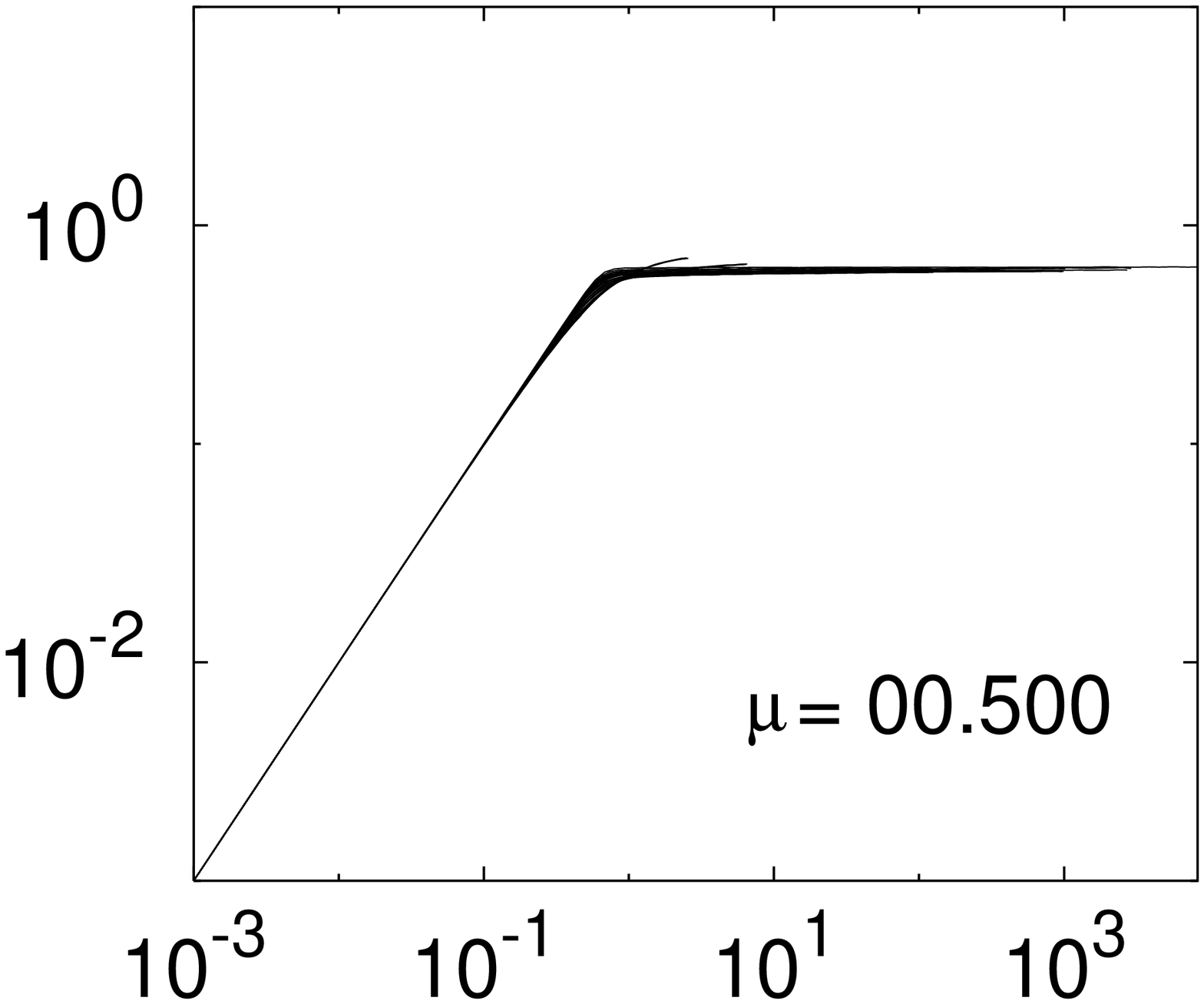,width=5.3cm,angle=270}
}\leftline{\hskip -1cm
\psfig{figure=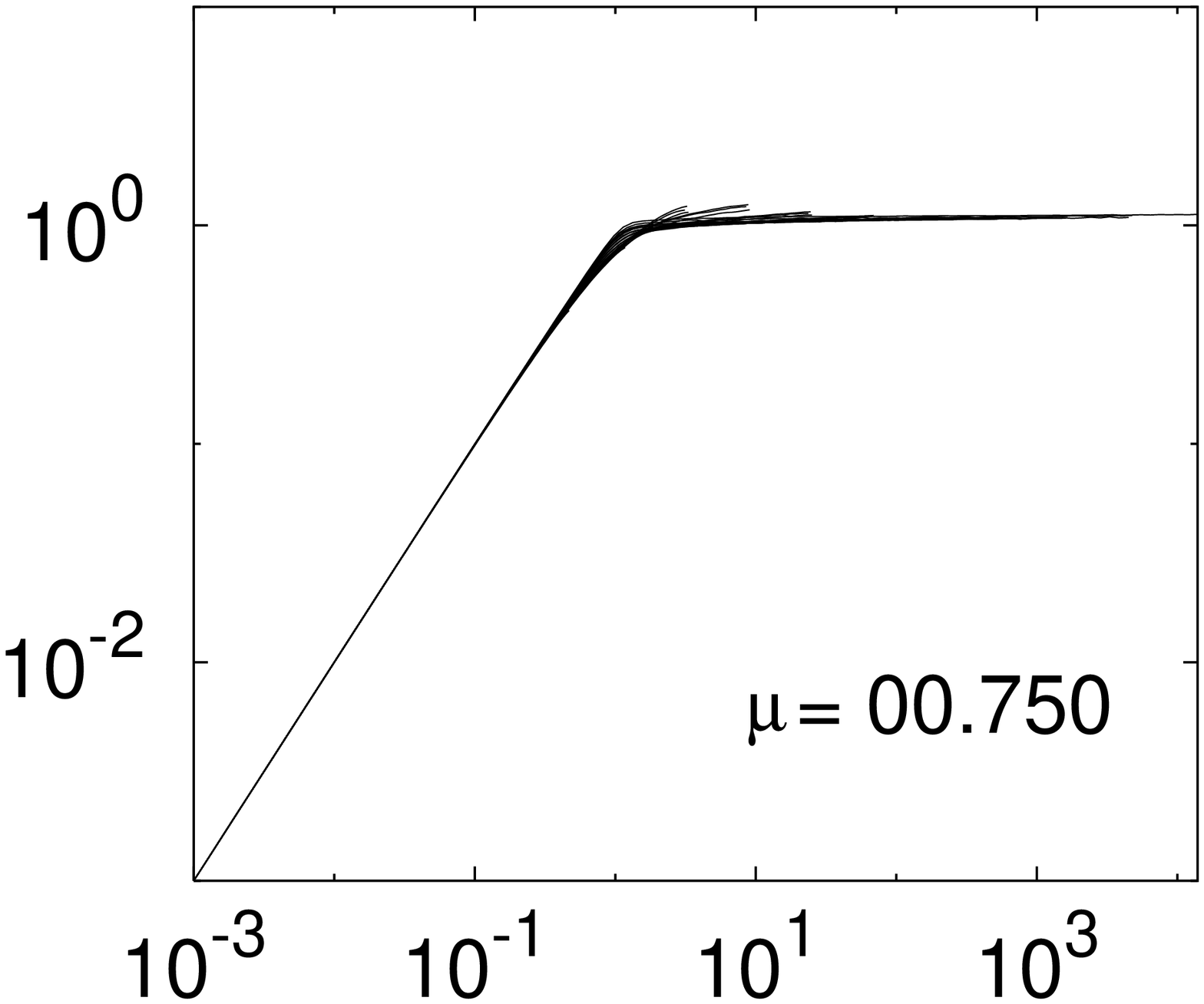,width=5.3cm,angle=270}\hskip -1cm
\psfig{figure=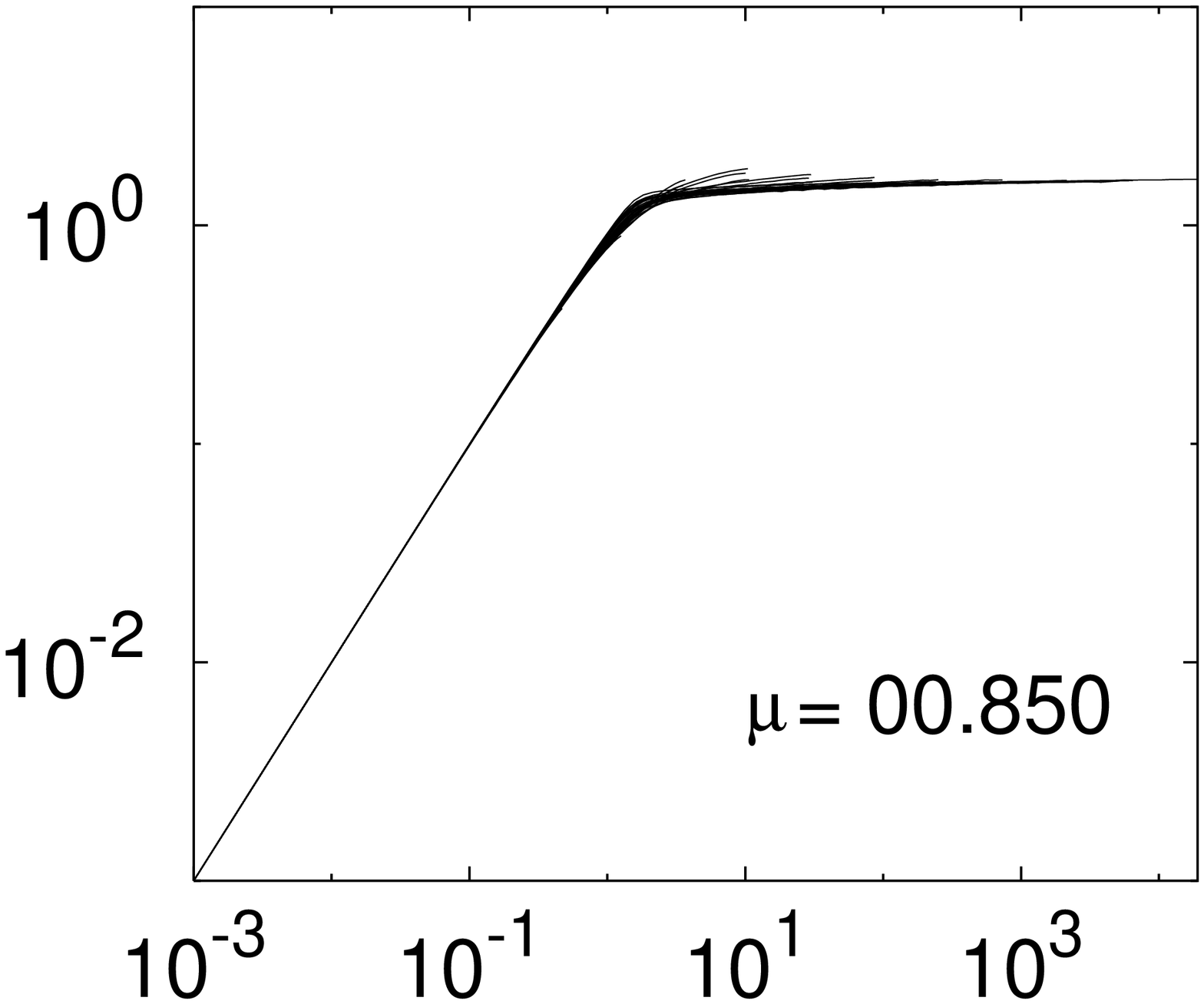,width=5.3cm,angle=270}
}\leftline{\hskip -1cm
\psfig{figure=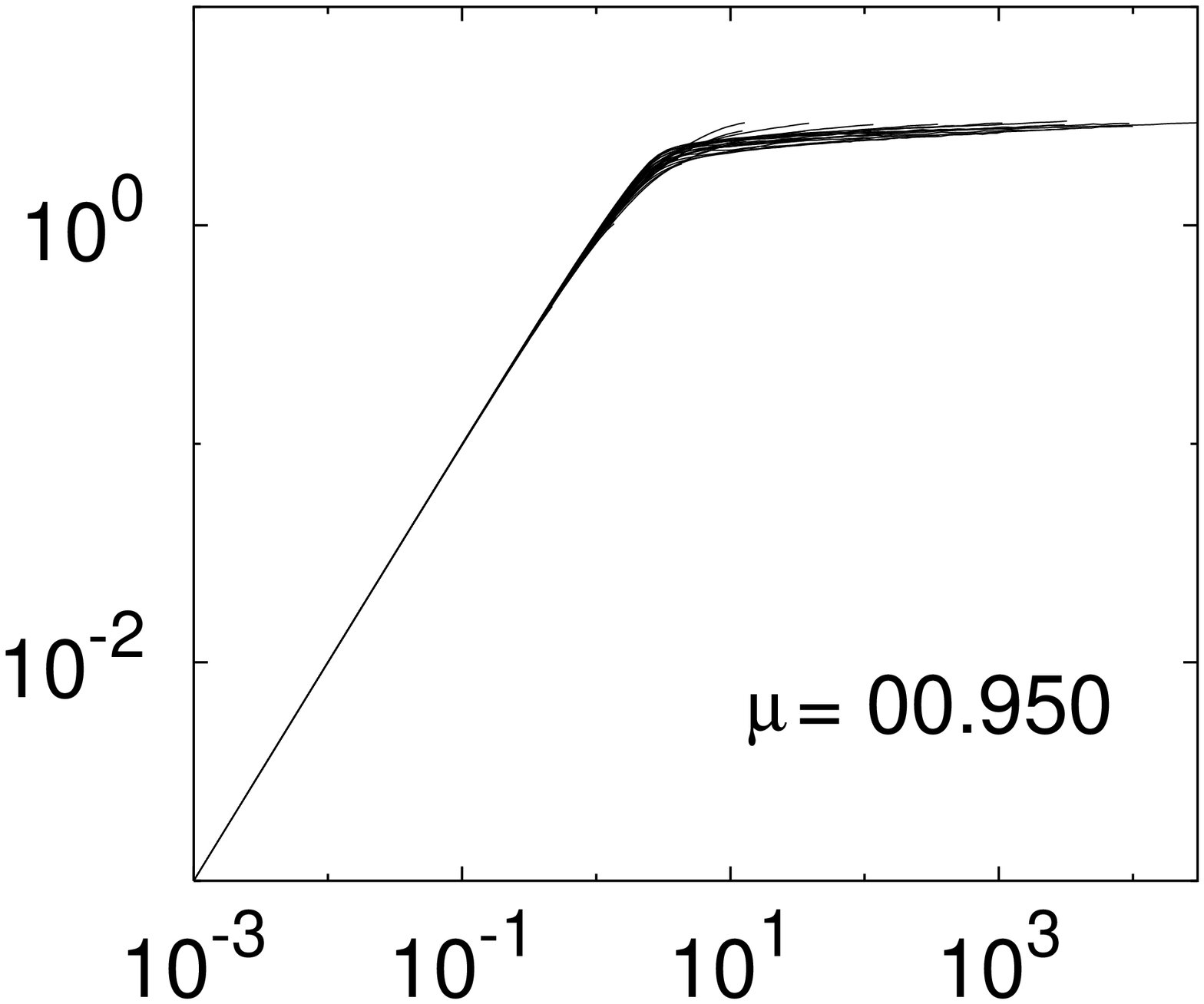,width=5.3cm,angle=270}\hskip -1cm
\psfig{figure=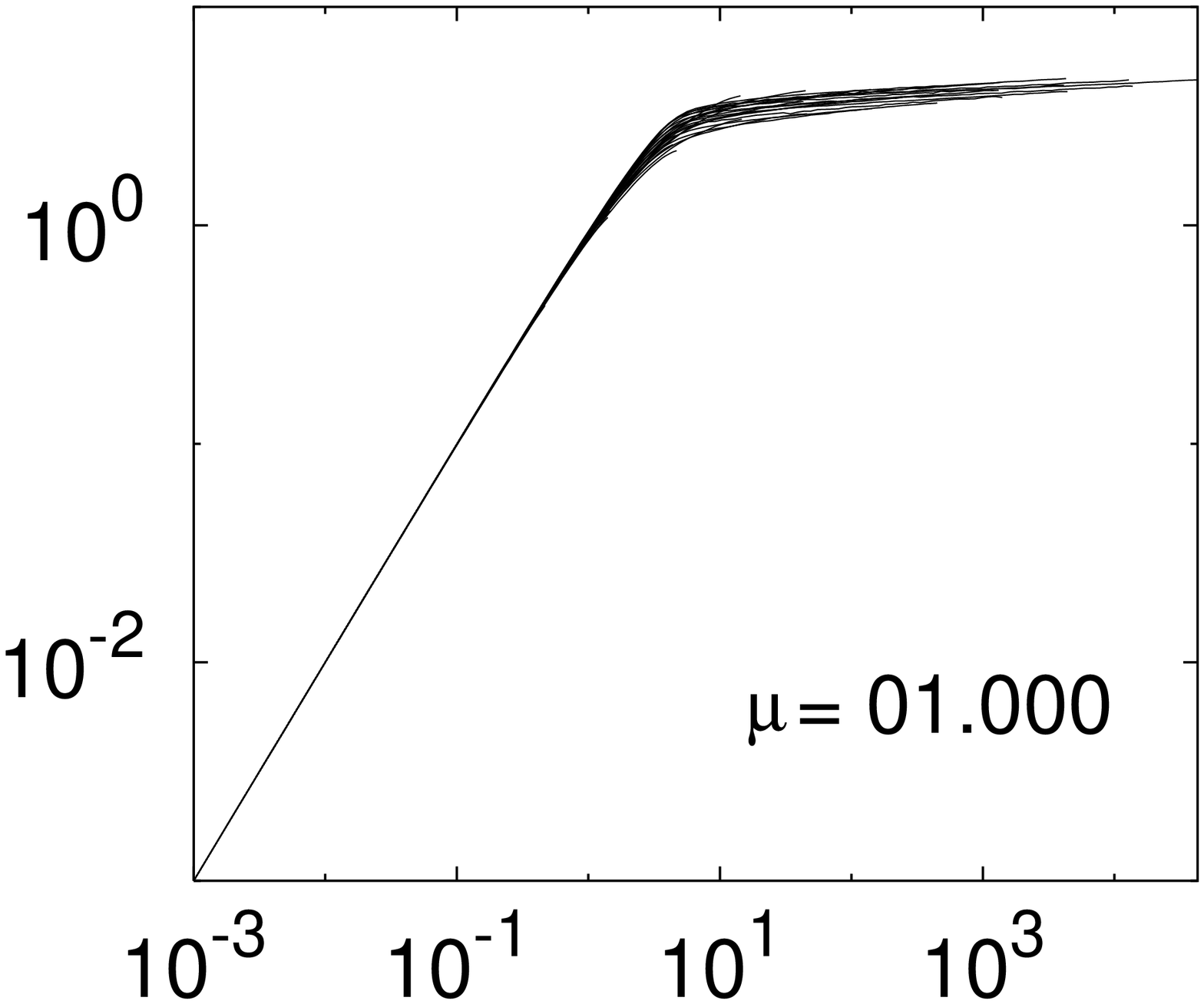,width=5.3cm,angle=270}
}
\caption{{} Shown is  $\ell/r_c$ vs $r/r_c$, with
  \hbox{$r_c=p^{-1}\log(4(pL)^{(1-\xpn)})/\log(4)$}. The density of LR bonds
  is $p=0.001,0.003$ and $0.01$. System sizes $L$ are of the form
  $L_k=10^{3+k/2}$ with $k=0,1\ldots,8$.  The proposed expression for $r_c$
  has only been justified theoretically for $\xpn=0$~\cite{MS99,D-MMPF00}, and
  is purely empirical for $0<\xpn<1$.}
\label{fig:ellofr_2}
}\end{figure}

The data in \Fig{fig:ell_0a} clearly show that $\ell(r)$ depends on system
size $L$ if $0 \leq \xpn < 1$. In the specific case $\xpn=0$, each of the
$L^{2d}/2$ possible LR bonds is present with the same probability $pL^{-d}$.
This corresponds to a $d$-dimensional lattice supplemented with $pL^d$ LR
bonds whose ends are randomly chosen, and goes under the name of Small-World
(SW)
network~\cite{WSC98,LS98,BAS99b,NWS99,MD99,MS99,D-MMPF00,MNE00,KAS01,MMPG01}.
In particular it was recently found~\cite{MS99,D-MMPF00} that on SW networks
($\xpn=0$) there is still a single characteristic length $r_c$ dictating the
behavior of $\ell(r)$, but it depends both on $L$ and $p$, and diverges as $L
\to \infty$, in any dimension $d$.  Analytic calculations~\cite{MS99}
confirmed by numerical measurement~\cite{D-MMPF00,MMPG01} show that, in $d$
dimensions,
\begin{equation}
\ell(r) = \left \{
\begin{array}{lcr}
r   & \hbox{for} & r < r_c
\\ \\
r_c & \hbox{for} & r > r_c,
\end{array}
\right .
\label{eq:ell_sw}
\end{equation}
where $r_c \sim p^{-1/d} \log(KpL^d)$ with $K$ a constant.
\\
In the particular case $d=1$ one has $r_c(\xpn=0,L,p)\sim p^{-1}\log(4pL)$. So
the $\xpn=0$ case is relatively simple, with $\ell(r)=r$ for $r<\log(4pL)/p$
and $\ell(r)=\log(4pL)/p$ for large $r$.
\\
By inspection of \Fig{fig:ell_0a} one concludes that $r_c$ depends on $L$ as
well as on $p,\xpn$ in the whole $0\leq \xpn < 1$ range.  When $\xpn=1$
however, the characteristic length dictating the behavior of $\ell(r)$ is
$\xi=p^{-1}$, and no longer $L$-dependent, as shown in
\Sec{sec:numerical1<a<2}. Guided by these observations, we now propose an
empirical expression for $r_c$ in terms of $L,p,\xpn$ in the whole $0\leq \xpn
\leq 1$ range.  This expression has to result in $\xi = p^{-1}$ when $\xpn \to
1$, and $r_c\sim p^{-1} \log(pL)$ when $\xpn \to 0$. It is easy to verify that
\begin{equation}
  r_c(\xpn,p,L) = p^{-1} \log 
\left [ 4(pL)^{(1-\xpn)}
\right ]/ \log(4)
\label{eq:rc_empirical}
\end{equation}
satisfies both requirements.  We find that this empirical expression gives
acceptable results for small $p$. In \Fig{fig:ellofr_2} we show $\ell(r)/r_c$
versus $r/r_c$ with $r_c$ given by (\Eqn{eq:rc_empirical}), for all values of
$L$ ranging from $10^3$ to $10^7$ and $p=0.001, 0.003$ and $0.010$. The
acceptable collapse of all data supports the validity of
(\Eqn{eq:rc_empirical}) reasonably well. \\
We find that $\ell(L)$ grows asymptotically as $\log(L)$ for $0<\xpn\leq 1$.
The naive path model already predicts a logarithmic behavior at $\xpn=1$ as
the following shows.  For $\xpn=1$ one has (see the beginning of
\Sec{sec:numerical}) $P_l=\log((l+1)/l)/\log(L)$, from which $G(x) \approx
x/\log(L)$. Thus \Eqn{eq:naive_formal} can be written approximately as
\begin{equation}
\ell_n^{(\xpn=1)}(r)=1 + \int_1^r \frac{dx}{1+ x/r_c},
\label{eq:log}
\end{equation}
where $r_c=p^{-1} \log(L)$. Thus naive-paths are determined, in the $\xpn \to
1$ limit, by a logarithmically $L$-dependent characteristic size $r_c$ and a
logarithmic behavior $\ell(r) \sim \log(r)$ above $r_c$. Given that actual
shortest-paths must be shorter than naive-paths, we conclude that $\ell(r)$ is
logarithmic for all $\xpn<1$.
\section{CONCLUSIONS}
\label{sec:conclusions}
We considered shortest paths on $d$-dimensional lattices of $L^d$ sites
supplemented with $pL^d$ long-range connections whose lengths $l$ are random
variables with power-law distribution $P(l) \sim l^{-\xpn}$. We call these
decaying probability (DP) networks, since it is the probability to have a LR
bond of length $l$, and not its strength, what decays with distance. The limit
$\xpn \to 0$ is the ``small-world'' network of Watts and
Strogatz~\cite{WSC98}. Under a rescaling transformation with scale parameter
$b$ in $d$ dimensions, a small local density $p$ of LR bonds transforms as
$\tilde p = b^{2d-\xpn} p$. In the $(\xpn,p)$ plane, $p=0$ is a repulsive
fixed line for $\xpn<2d$ and an attractive fixed line for $\xpn>2d$. Thus
rescaling arguments predict $\xpn_c=2d$ to be a critical decay rate above
which LR bonds are irrelevant.  Particularizing to $d=1$, a directed model
that gives an upper bound for shortest-paths can be analytically solved
(\Sec{sec:naive}) and has three regions in the $\xpn$-axis: a) $\ell(r)
\propto r$ for $\xpn>2$, b) $\ell(r) \propto r^{\xpn-1}$ for $1<\xpn<2$ and c)
$\ell(r)$ logarithmic for $\xpn<1$.  In accordance with rescaling arguments,
we find numerically that in one dimension $\xpn=2$ is a critical point
separating a ``short-range phase'' ($\xpn>2$) where shortest-path lengths are
linear, $\ell(r) \propto r$, from a ``long-range phase'' ($\xpn<2$) where
shortest-path lengths are sublinear, $\ell(r) \propto r^{\theta_s}$ with
$\theta_s<1$. Our finding that $\xpn_c=2$ for all $p$ is consistent with
previous work of Jespersen and Blumen~\cite{JBS00}, but is in disagreement
with recent claims of Sen and Chakrabarty~\cite{SCS01} who suggest the
existence of a $p$-dependent boundary $\xpn^{*}(p)$. We showed that this
apparent boundary is a finite-size effect, due to the fast growth of a
correlation length $\xi$ as $\xpn \to 2^-$.

For small $p$ and $1\leq \xpn \leq 2$, a characteristic size $\xi = p^{-\nu}$
with $\nu= 1/(1-\xpn)$ dictates the shortest path properties. For $r<\xi$ one
has $\ell(r) \approx r$ while for $r>> \xi$, $\ell(r) \sim r^\theta_s(\xpn)$
is found. This divergence in the correlation length exponent $\nu$ as $\xpn
\to 2^-$ is of the same kind as reported for spin models
previously~\cite{KP76,CMT97,LBC97}.

For $\xpn <1$ the characteristic size behaves as $p^{-1}$ but is also
$L$-dependent and we find that \Eqn{eq:rc_empirical} provides a good empirical
fit of both its $p$- and $L$-dependence.

The asymptotic exponent $\theta_s$ is found numerically to attain its
short-range value $\theta_s=1$ for $\xpn>2$. It is discontinuous at $\xpn=2$,
where it probably takes a value near $1/2$, and then goes to zero smoothly as
$\xpn \to 1^+$. For $\xpn \leq 1$ we find logarithmic (or Mean Field)
behavior: $\theta_s=0$ and $\ell(r) \sim \log(r)$ asymptotically. For $\xpn \to
0$ $\ell(r)$ saturates at large distances to a value which depends
logarithmically on system size~\cite{MS99,D-MMPF00,MMPG01}.
\section{Acknowledgments}
The authors acknowledge finantial support of FAPERJ, CNPq and CAPES (Brazil)
and CONACYT (Mexico).
\appendix
\section{Shortcut distribution}
\subsection{Normalization}
The scale-invariant shortcut distribution $p(r)$
\Eqn{eq:invariant-distribution} can be approximated by $p(r) \approx 1$ for
$r<r_c=\invdens^{1/\xpn}$ and $p(r) \approx \invdens/r^\xpn$ for $r>r_c$. Thus
the normalization condition \Eqn{eq:normalization} can be written as
\begin{equation}
p = S_d \left \{ 
\begin{array}{lr}
\int_1^{r_c} r^{d-1}dr + \invdens \int_{r_c}^L r^{d-1-\xpn}dr 
& \hbox{for~} r_c >1
\\ \\
\invdens \int_{1}^L r^{d-1-\xpn}dr & \hbox{for~} r_c < 1,
\end{array}
\right . 
\end{equation}
so that if $V_d=S_d/d$ is the volume of a unit radius sphere,
\begin{equation}
p = V_d \left \{ 
\begin{array}{lr}
\frac{\invdens d L^{d-\xpn}}{d-\xpn} - \frac{\xpn \invdens^{d/\xpn}}{d-\xpn}-1
& \hbox{for~} \invdens > 1
\\ \\
\frac{\invdens d (L^{d-\xpn}-1)}{d-\xpn} 
& \hbox{for~} \invdens < 1
\end{array}
\right . 
\end{equation}
When $\xpn<d$ and if $p$ remains finite in the $L\to \infty$ limit one has
that
\begin{equation}
\label{eq:rel1}
\invdens = p \frac{d-\xpn}{S_d} L^{-(d-\xpn)}
\end{equation}
This goes to zero for large $L$, which justifies the power-law approximation
\Eqn{eq:pl-approx}. For $\xpn>d$ on the other hand, and assuming $p$ small,
\begin{equation}
\label{eq:rel2}
\invdens = p \frac{\xpn-d}{S_d},
\end{equation}
so that the power-law approximation holds for any finite $p$ when $\xpn<d$ but
only for $p$ small when $\xpn>d$. The power-law distribution is properly 
normalized when
\begin{equation}
1=C \int_1^L r^{d-1-\xpn} dr \quad \rightarrow \quad
C = \frac{(\xpn-d)  L^{\xpn-d}}{S_d (L^{\xpn-d}-1)}
\end{equation}
therefore in the limit of large $L$,
\begin{equation}
p(r) =
\left \{
\begin{array}{lr}
\frac{(\xpn-d)}{S_d} \quad \frac{p}{r^{\xpn}} & \hbox{for~} \xpn > d
\\
\\
\frac{(d-\xpn)}{S_d L^{d-\xpn}} \quad \frac{p}{r^{\xpn}}& \hbox{for~} \xpn < d
\end{array}
\right .
\label{eq:pofl}
\end{equation}
gives the probability for two sites separated by an Euclidean distance $r$ to
be connected by a LR bond. 
\subsection{Rescaling}
\label{apd:rescaling}
From the rescaling of $\invdens$, \Eqn{eq:rescaling}, and the relationships
(\ref{eq:rel1}) and (\ref{eq:rel2}) between $\invdens$ and $p$ it is immediate
to conclude that,
\begin{equation}
\tilde p = b^{y_p} p,
\end{equation}
with
\begin{equation}
y_p = \left \{
\begin{array}{crr}
d & \hbox{for} & \xpn \leq d \\ \\
2d-\xpn &\hbox{for} & \xpn > d
\end{array}
\right .
\end{equation}
\subsection{Naive paths when $\xpn>2$}
\label{apd:naive}
For $\xpn>1$ and $L>>1$ we have that $P_l=l^{1-\xpn}-(l+1)^{1-\xpn}$.  Using
this expression, \Eqn{eq:G} gives $G(x)=\sum_{l=1}^x (l-1)\left
  [l^{1-\xpn}-(l+1)^{1-\xpn} \right ]$
$=H(x+1,\xpn-1)-1-(x+1)^{-\xpn}+(x+1)^{-(\xpn-1)}$, where
$H(x,\alpha)=\sum_{l=1}^x 1/l^\alpha$ are called Harmonic Numbers.
$H(x,\alpha)$ can be approximated (within one percent error) for all $\alpha
\geq 1$ and $x\geq 2$ by
\begin{equation}
H(x,\alpha) \approx 1 + 2^{-\alpha} + \frac{3^{-\alpha}+x^{-\alpha}}{2}
+ \frac{x^{1-\alpha}-3^{1-\alpha}}{1-\alpha}
\end{equation}
Within this approximation one obtains $(\bar l-1)=H(\infty,\xpn-1)-1=
2^{1-\xpn} + \frac{3^{1-\xpn}}{2} + \frac{3^{2-\xpn}}{\xpn-2}$, which is found
to be very accurate for all $\xpn \geq 2$. Using this approximate expression,
\Eqn{eq:phi} can be integrated to give
\begin{eqnarray}
\Phi(r)&=& 
\frac{(\bar l-1)(r-1) }{r}
-\frac{Z(r,3-\xpn)}{r(\xpn-2)}
\cr \cr &+& \frac{3 Z(r,2-\xpn)}{2r}
- \frac{Z(r,1-\xpn)}{r}
\label{eq:phi_approx}
\end{eqnarray}
where $Z(x,\alpha)=((x+1)^{\alpha}-2^{\alpha})/\alpha$. 

\bibliographystyle{apsrev}

\bibliography{SMALL-WORLD,LONG-RANGE,Books}
\end{document}